 \documentclass[final,5p,times,twocolumn]{elsarticle}

\usepackage{amssymb}
\usepackage{amsthm}
\usepackage{amsmath}
\usepackage{siunitx}

\journal{~}

\begin{document}

\begin{frontmatter}



\title{Global turbulence simulations of the tokamak edge region with GRILLIX}

\author[label1]{A.~Stegmeir}
\ead{Andreas.Stegmeir@ipp.mpg.de}
\author[label1]{A.~Ross}
\author[label1]{T.~Body}
\author[label2,label3]{M.~Francisquez}
\author[label1]{W.~Zholobenko}
\author[label1]{D.~Coster}
\author[label1]{O.~Maj}
\author[label1]{P.~Manz}
\author[label1]{F.~Jenko}
\author[label3]{B.N.~Rogers}
\author[label1]{K.S.~Kang}

\address[label1]{Max-Planck-Institut f\"{u}r Plasmaphysik, D-85748 Garching, Germany}
\address[label2]{MIT Plasma Science and Fusion Center, Cambridge, MA 02139, USA}
\address[label3]{Department of Physics and Astronomy, 6127 Wilder Laboratory, Dartmouth College, Hanover, NH 03755, USA}

\begin{abstract}
Turbulent dynamics in the scrape-off layer (SOL) of magnetic fusion devices is intermittent with large fluctuations in density and pressure. Therefore, a model is required that allows perturbations of similar or even larger magnitude to the time-averaged background value. The fluid-turbulence code GRILLIX is extended to such a global model, which consistently accounts for large variation in plasma parameters. Derived from the drift reduced Braginskii equations, the new GRILLIX model includes electromagnetic and electron-thermal dynamics, retains global parametric dependencies and the Boussinesq approximation is not applied. The penalisation technique is combined with the flux-coordinate independent (FCI) approach [F.~Hariri and M.~Ottaviani, Comput.~Phys.~Commun.~{\bf 184}:2419, (2013); A.~Stegmeir et al., Comput.~Phys.~Commun.~{\bf 198}:139, (2016)], which allows to study realistic diverted geometries with X-point(s) and general boundary contours. We characterise results from turbulence simulations and investigate the effect of geometry by comparing simulations in circular geometry with toroidal limiter against realistic diverted geometry at otherwise comparable parameters. Turbulence is found to be intermittent with relative fluctuation levels of up to $40\%$ showing that a global description is indeed important. At the same time via direct comparison, we find that the Boussinesq approximation has only a small quantitative impact in a turbulent environment. In comparison to circular geometry the fluctuations are reduced in diverted geometry, which is related to a different zonal flow structure. Moreover, the fluctuation level has a more complex spatial distribution in diverted geometry. Due to local magnetic shear, which differs fundamentally in circular and diverted geometry, turbulent structures become strongly distorted in the perpendicular direction and are eventually damped away towards the X-point.
\end{abstract}


\end{frontmatter}


\section{Introduction}

Understanding the complex multi-physics of the edge region -- the scrape-off layer (SOL) and closed-field line region immediately near the separatrix -- is of critical importance for the development of fusion energy. Due to the relative stiffness of the internal profiles of temperature and density, the core values and therefore the overall fusion performance is strongly determined by the edge profiles. Furthermore, in a fusion reactor there will be a large exhaust of particles and heat due to imperfect confinement. This plasma exhaust is directed towards divertor target plates, and preventing these fluxes from exceeding engineering limits, above which the performance and lifetime of the reactor is significantly reduced, is a high-priority area of fusion research. Prediction of these heat fluxes for future devices such as ITER or DEMO is complicated by uncertainty in extrapolation of the width of the exhaust channel from current devices \cite{eich:solwidth13,halpern:solwidth13}.

Modelling of the edge plasma is a significant challenge due to the highly-coupled interplay of multiple different physics regimes and disparate spatial and temporal scales. Magnetized plasma physics, complex magnetic geometry, neutral physics and momentum transfer, atomic and molecular chemistry, radiation from excited states, wall recombination, surface chemistry and impurity sputtering all can affect the edge plasma. Furthermore, the edge can exhibit phenomena over a large range of spatial and temporal scales -- from the formation of small-scale intermittent turbulent filaments to large-scale long-timescale effects such as equilibration of the background in response to the magnetic and wall geometry. Inclusion of an extended physics set or finer spatial and temporal scales typically improves the accuracy of the code with respect to experiment, but at the expense of increased computational cost. Within the subset of codes based on the multi-fluid approximation, the two broad classes are `transport' and `turbulence' codes. Transport codes such as SOLPS code \cite{wiesen:solps15} include a significant range of multi-physics but do not treat turbulent transport self-consistently. Instead, the effects of turbulence are approximated via an effective diffusion, which remains an ad-hoc input. In contrast, turbulence codes self-consistently treat turbulence by evolving the 3D Braginskii models \cite{braginskii65} -- at the cost of increased runtime and/or a reduced physics set.

Several recent projects aim at developing fluid-turbulence codes, of which we note the GBS \cite{ricci:gbs12,halpern:gbs16,paruta:gbsx18}, HERMES (BOUT++) \cite{dudson:hermes17}, TOKAM3X \cite{tamain:tokam3x16}, GDB \cite{zhu:gdb18} and GRILLIX \cite{stegmeir:ppcf18} projects. In contrast to the other codes mentioned, the GRILLIX project is notable for its use of the flux-coordinate independent (FCI) approach \cite{hariri:fenicia13,stegmeir:cpc16,stegmeir:fciaddendum17}. This method prevents the issue of coordinate singularities at the separatrix and X-point which arise from the use of field- or flux-aligned coordinates. GRILLIX employs a cylindric grid $(R_i,\varphi_k,Z_j)$ where parallel operators are discretised via field line tracing between toroidal planes and field line map interpolation within each plane. This allows for the use of a single consistent method to be used for all grid points, including the possibility of (possibly multiple) X-points. Furthermore, the use of Cartesian grids prevents resolution imbalances between the outboard mid plane and the X-point region, allowing the dynamics around the X-point to be investigated with high accuracy. To allow for the treatment of general non-conformal boundaries, the penalization method is used to enforce the desired boundary conditions.

In this paper, the extension of GRILLIX by electromagnetic and electron thermal dynamics is presented, with the resulting model being a global drift reduced Braginskii model. Here 'global' means that parametric dependencies are kept and that the Boussinesq approximation is not applied, i.e.~nowhere a separation is made between fluctuations and background for the density and temperature. This is needed for the consistent description of high amplitude fluctuations which are regularly observed in experiments \cite{wootton:edgesolcharacter90}. Recently several codes were adapted to abolish the Boussinesq approximation also for turbulence applications \cite{halpern:gbs16,zhu:gdb18}, whereat now the geometrical complexity of diverted equilibria is additionally introduced in GRILLIX. The Boussinesq approximation has often been studied at isolated sub problems, i.e.~blob propagation \cite{angus:nonbsq14,militello:blobs17,ross:phd18,ross:bossinesq18}. Here we investigate its impact in a fully turbulent environment and find that it has only minor quantitative impact on results. Using parameters characteristic for the COMPASS tokamak \cite{panek:compass15} turbulence simulations are carried out with GRILLIX. The turbulence features  intermittency and exhibits large relative fluctuation levels, which shows that a global description is generally important. The impact of geometry is studied by comparing simulations in circular geometry with toroidal limiter against realistic diverted geometry at otherwise comparable parameters. The fluctuation level is reduced in diverted geometry owed to a different zonal flow structure. Moreover, the fluctuation level exhibits a more complex spatial distribution, which we explain as a consequence of local magnetic shear, which fundamentally differs in the edge region between circular and diverted geometry. Strong local magnetic shear causes a distortion of turbulent structures, which become subsequently subject to strong perpendicular dissipation. The X-point, where magnetic shear becomes locally very strong, thereby tends to disconnect the low field side from the high field side, where curvature acts as stabilizing \cite{farina:xpoint93}. Therefore, stronger poloidal asymmetries in the fluctuation level are observed in diverted geometry.

The remainder of this paper is organized as follows: In section \ref{sec:Physical_model} we present the physical model employed in GRILLIX, a global 3D drift reduced Braginskii model \cite{zeiler:drift_approx97}. With respect to the previous version \cite{stegmeir:ppcf18} GRILLIX has been extended by electromagnetic dynamics, electron temperature dynamics and the Boussinesq approximation has been relaxed, which enables to simulate plasma turbulence globally, i.e.~without splitting quantities into background and fluctuations. The implementation of important new features is described in section \ref{sec:GRILLIX}. A geometric multigrid solver for the generalised 2D perpendicular Helmholtz equation allows an efficient treatment of the new electromagnetic terms and relaxation of the Boussinesq approximation. The treatment of sheath boundary conditions at the divertor/limiter plates via penalization techniques was motivated from the GDB code \cite{zhu:gdb18,francisquez:phd18} and we give a generalisation to diverted geometries. The extended model and new features are verified by analytic means and the method of manufactured solutions (MMS) \cite{salari:mms00}. In section \ref{sec:Simulation_results} we characterize edge turbulence with GRILLIX simulations and clarify the impact of geometry by comparing simulations in circular geometry with toroidal limiter against simulations in diverted geometry at otherwise comparable parameters. A summary and outlook is given in section \ref{sec:Conclusions_and_outlook}.

\section{Physical model}\label{sec:Physical_model}
\subsection{Global drift reduced Braginskii equations}
Based on the assumptions of short mean free paths, i.e.~$\lambda_c\ll R_0$, the drift reduced Braginskii model describes plasma dynamics of low frequency ($\omega\ll\Omega_i$) in comparison to the ion cyclotron frequency $\Omega_i$ \cite{braginskii65,zeiler:drift_approx97}.It is suitable to describe turbulence at low temperature in the edge region self consistently. As further practical approximations cold ions ($T_i\ll T_e$) are assumed in GRILLIX, and whereas electromagnetic effects are kept in Ohm's law, magnetic flutter is neglected, i.e.~transport is assumed to be electrostatic and the description of certain phenomena like Edge Localised Modes (ELM's) is excluded. 

The following normalisation is employed: Time $t$ is normalised against $R_0/c_{s0}$ with the sound speed $c_{s0}:=\sqrt{T_{0}/M_i}$ at some reference temperature $T_{0}$. Parallel scales $x_{\parallel}$ are normalised against major radius $R_0$ and perpendicular scales $x_{\perp}$ against the sound Larmor radius $\rho_{s0}:=c\sqrt{T_0M_i}/(eB_0)$ with $B_0$ the magnetic field strength on axis and $M_i$ is the ion mass. The dynamical variables in GRILLIX are density $n$ normalized against some reference density $n_0$, electron temperature $T_e$ against $T_0$, parallel ion $u_\parallel$ and electron $v_{\parallel}$ velocities against $c_{s0}$, parallel current $j_\parallel$ against $en_0c_{s0}$, electrostatic potential $\phi$ against $T_0/e$ and the parallel component of the electromagnetic potential $A_{\parallel}$ against $\beta_0B_0\rho_{s0}$ with $\beta_0:=4\pi n_0T_0/B_0^2$ the dynamical plasma beta at reference values. In order to preserve its positivity the logarithms of normalised density $\theta_n:=\log n$ and temperature $\xi_e:=\log T_e$ are evolved in time. Finally, the normalised set of equations implemented in GRILLIX are:
\begin{align}
\frac{d_u}{dt}\theta_n+\nabla\cdot\left(\mathbf{b}u_\parallel\right)=&\mathcal{C}(\phi)-T_e\left[\mathcal{C}(\xi_e+\theta_n)\right]+\frac{1}{n}\nabla\cdot\left(\mathbf{b}j_\parallel\right) \notag\\
&+\frac{1}{n}\left[\mathcal{D}_n\left(n\right)+S_n\right],
\label{eq:continuity}
\end{align}

\begin{align}
\nabla\cdot\left[\frac{n}{B^2}\frac{d_u}{dt}\nabla_\perp\phi\right]=&-nT_e\left[\mathcal{C}(\theta_n+\xi_e)\right]+\nabla\cdot\left(\mathbf{b}j_\parallel\right) \notag \\
 & + \mathcal{D}_w\left(\Omega\right),
\label{eq:vorticity}
\end{align}

\begin{align}
\frac{d_u}{dt}u_\parallel=&-T_e\left[\nabla_\parallel\theta_n+\nabla_\parallel\xi_e\right]+\mathcal{D}_u(u_\parallel),
\label{eq:parmomentum}
\end{align}

\begin{align}
\frac{\partial}{\partial t}\psi_\parallel + \mu\left(\mathbf{v}_E\cdot\nabla+v_\parallel\nabla_\parallel\right)\left(\frac{j_\parallel}{n}\right)=&-\left(\frac{\eta_{\parallel 0}}{T^{3/2}}\right)j_\parallel \notag \\
& +T_e\left[\nabla_\parallel\theta_n+1.71\nabla_\parallel\xi_e\right]-\nabla_\parallel\phi, \notag \\
& + \mathcal{D}_p\left(\psi_\parallel\right)
\label{eq:ohm}
\end{align}

\begin{align}
\frac{d_v}{dt}\xi_e=&\frac{2}{3}\mathcal{C}(\phi)-T_e\left[\frac{2}{3}\mathcal{C}(\theta_n)+\frac{7}{3}\mathcal{C}(\xi_e)\right]-\frac{2}{3}\nabla\cdot\left(\mathbf{b}v_\parallel\right) \notag \\ 
+\frac{2}{3}0.71\frac{1}{n}\nabla\cdot\left(\mathbf{b}j_\parallel\right) 
&\frac{2}{3}\left(e^{-\frac{5}{2}\xi_e-\theta_n}\eta_{\parallel0}\right)j_\parallel^2+\frac{2}{3}e^{-\theta_n-\xi_e}\nabla\cdot\left[\chi_{\parallel 0}e^{\frac{7}{2}\xi_e}\nabla_\parallel\xi_e\right] \notag \\
&\frac{1}{T_e}\left[\mathcal{D}_t\left(T_e\right)+S_t\right],
\label{eq:etemperature}
\end{align}

\begin{align}
\nabla_\perp^2A_\parallel=-j_\parallel.
\label{eq:faraday}
\end{align}
Equations (\ref{eq:continuity}-\ref{eq:faraday}) are the electron continuity equation, vorticity equation or quasineutrality condition, parallel momentum equation, Ohm's law, electron temperature equation and Faraday's law respectively. The total time derivatives contain advection by the E$\times$B drift and the parallel velocity, i.e.~$\frac{d_u}{dt}:=\frac{\partial}{\partial t}+\mathbf{u}_E\cdot\nabla+u_\parallel\nabla_\parallel$ and $\frac{d_v}{dt}:=\frac{\partial}{\partial t}+\mathbf{u}_E\cdot\nabla+v_\parallel\nabla_\parallel$. The parallel gradient $\nabla_\parallel=\mathbf{b}\cdot\nabla$ is taken along the static equilibrium magnetic field $\mathbf{B}$ with unit vector $\mathbf{b}:=\mathbf{B}/B$, i.e.~magnetic flutter is not included. As auxiliary variables the generalised vorticity $\Omega:=\nabla\cdot\left(\frac{n}{B^2}\nabla_\perp\phi\right)$ and the generalised parallel electromagnetic potential $\psi_\parallel:=\beta_0A_\parallel +\mu\frac{j_\parallel}{n}$ have been introduced. Within the FCI approach (see section \ref{sec:Spatial_discretisation}) the advection by E$\times$B velocity and the curvature operator can be written as:
\begin{align*}
\mathbf{u}_E\cdot\nabla f=& \frac{\delta}{B^2}\left(\mathbf{B}\times\nabla \phi\right)\cdot\nabla f\approx -\frac{\delta}{B}\left[\phi,f\right]_{R,Z}, \\
\mathcal{C}(f)=&-\delta\left[\left(\nabla\times\frac{\mathbf{B}}{B^2}\right)\cdot\nabla f\right]\approx -2\partial_Zf,
\end{align*}
with the Jacobi bracket $\left[\phi,f\right]_{R,Z}:=\partial_R\phi\partial_Zf-\partial_Z\phi\partial_Rf$. The dimensionless parameters of the system are the drift scale $\delta:=R_0/\rho_{s0}$, electron to ion mass ratio $\mu:=\frac{m_e}{M_i}$, the dynamical plasma beta $\beta_0$, reference parallel resistivity $\eta_{\parallel0}:=0.51\mu/ (\tau_{e0}(c_{s0}/R_0))$ with $\tau_{e0}$ the electron-ion collision time at reference values and the reference parallel heat conductivity $\chi_{\parallel0}:=3.15/\mu\tau_{e0}(c_{s0}/R_0)$. Particle and thermal source terms $S_n,S_t$ have been added, and mostly for numerical reasons also dissipation terms:
\begin{align}
\mathcal{D}_f:=\nu_{\perp,f}\nabla_\perp^{2N}f + \nu_{\parallel,f}\nabla\cdot\left(\mathbf{b}\nabla_\parallel f\right),
\end{align}
with constant coefficients $\nu_{\perp,f},\,\mu_{\parallel,f}$, and $N$ controlling the order of perpendicular dissipation. Note that in order to ensure conservation of particles the dissipation in the continuity equation (\ref{eq:continuity}) does not act on the logarithm of the density $\theta_n$, but the density $n$.

The model is global in the sense that no separation of variables between a background part and fluctuating part is made and the dependency of the parallel resistivity and heat conduction on density and temperature is kept. Moreover the dependency on the density in the polarization term of the vorticity equation (\ref{eq:vorticity}) is also kept, i.e.~the Boussinesq approximation is not applied. The model conserves energy apart from the fact that we have neglected advection by the polarization velocity, which has been shown to have only a minor effect on conservation of energy \cite{ross:nonbsq17}.

\subsection{Boundary conditions}
The simulation domain in GRILLIX is usually bounded by an inner (core) limiting flux surface, an outer (wall) limiting flux surface and depending on geometry by limiter or divertor baffle plates. 

Sheath physics determining the boundary conditions for the divertor/limiter is a rich topic by itself. A sophisticated set of boundary conditions is given e.g.~in \cite{loizu:bndconds12}, where an inclination of the magnetic field with respect to the target plates is taken into account, but more commonly just Bohm boundary conditions \cite{stangeby:plasmaboundary00} are employed. Moreover, the treatment of sheath boundaries within the FCI approach is also numerically cumbersome for which a penalization method is employed in GRILLIX described in section \ref{subsec:penalization}. Due to all these complexities we restrict ourselves firstly to the relatively simple but robust set of insulating sheath boundary conditions. This assumption implies, e.g.~that blobs propagate purely according to the inertial scaling $v_{blob}\propto w_{blob}^{1/2}$, with $v_{blob}$ the radial blob velocity and $w_{blob}$ the blob width, whereas with Bohm boundary conditions larger blobs would propagate slower according to the sheath connected scaling $v_{blob}\propto w_{blob}^{-2}$ \cite{krasheninnikov:blobs08}. Therefore, the assumption of an insulating sheath might appear strong, but is of relevance for modelling detached conditions where resistivity in the front of the target plates is strongly enhanced due to the presence of neutrals. 

Finally, the boundary conditions employed in GRILLIX are:
\begin{align}
&u_\parallel \gtrless\pm\sqrt{T_e}, \label{eq:sheathbnd_upar}\\ 
& j_\parallel = 0,\\
& \phi = \Lambda T_e \\
&-\chi_\parallel\nabla_\parallel T_e = \gamma_e T_e u_\parallel, \label{eq:sheathbnd_te} \\
& \nabla_\parallel n=0,
\end{align}
where the upper/lower sign denotes if the direction of magnetic field is directed towards/away from target plates, $\Lambda\approx0.5\ln\left(\frac{M_i}{2\pi m_e}\right)$ is the sheath floating potential and $\gamma_e\approx 2.5$ the effective electron sheath transmission factor. As the continuity equation is of hyperbolic nature in the parallel direction the boundary condition on the density should be as unrestricted as possible, and we found that a homogeneous Neumann boundary condition is numerically more robust than extrapolation $(\nabla_\parallel^2n=0)$. The generalised vorticity and electromagnetic potential are obtained consistently with the electrostatic potential respectively the parallel current as described in section \ref{subsec:penalization}.

In the radial direction either homogeneous Neumann boundary conditions (for $n$, $T_e$, $u_\parallel$) are applied or homogeneous Dirichlet boundary conditions (for $A_\parallel$, $j_\parallel$, $\Omega$). An exception is the electrostatic potential which is set at the wall to $\left.\phi\right|_{wall}=\Lambda T_e$. In order to avoid fluxes of energy and particles due to $E\times B$ drifts through the core, the potential has to be constant on the inner limiting flux surface, and we set it to $\phi_{core} = \Lambda\left\langle T_e\right\rangle_{LCFS}$, where $\left\langle T_e\right\rangle_{LCFS}$ is the zonal averaged electron temperature on the last closed flux surface. The motivation for this stems from the fact that due to the sheath boundary conditions the potential follows roughly $\phi\sim \Lambda T_e$ in the SOL, and the chosen boundary condition does therefore not allow a global radial electric field in the closed flux surface region. The large scale radial electric field in the closed field line region is determined by effects that are not yet included in the GRILLIX model, e.g.~ion pressure gradient \cite{stroth:er11,viezzer:er13}. Therefore, the self-consistent modelling of the global radial electric field in the closed field line region is postponed until at least ion temperature effects will be taken into account in GRILLIX. We want to note that from the geometrical point of view GRILLIX is able to deal easily with the full tokamak including the core region with O-point \cite{stegmeir:ppcf18}. Whereas this would exclude a possibly spurious influence from core boundaries and therefore allow a more self-consistent approach, the many additional grid points would pose a large computational overhead.

\section{GRILLIX}\label{sec:GRILLIX}
\subsection{Spatial discretisation}\label{sec:Spatial_discretisation}
\begin{figure}
\includegraphics[trim=150 100 50 0,clip,width=1.0\linewidth]{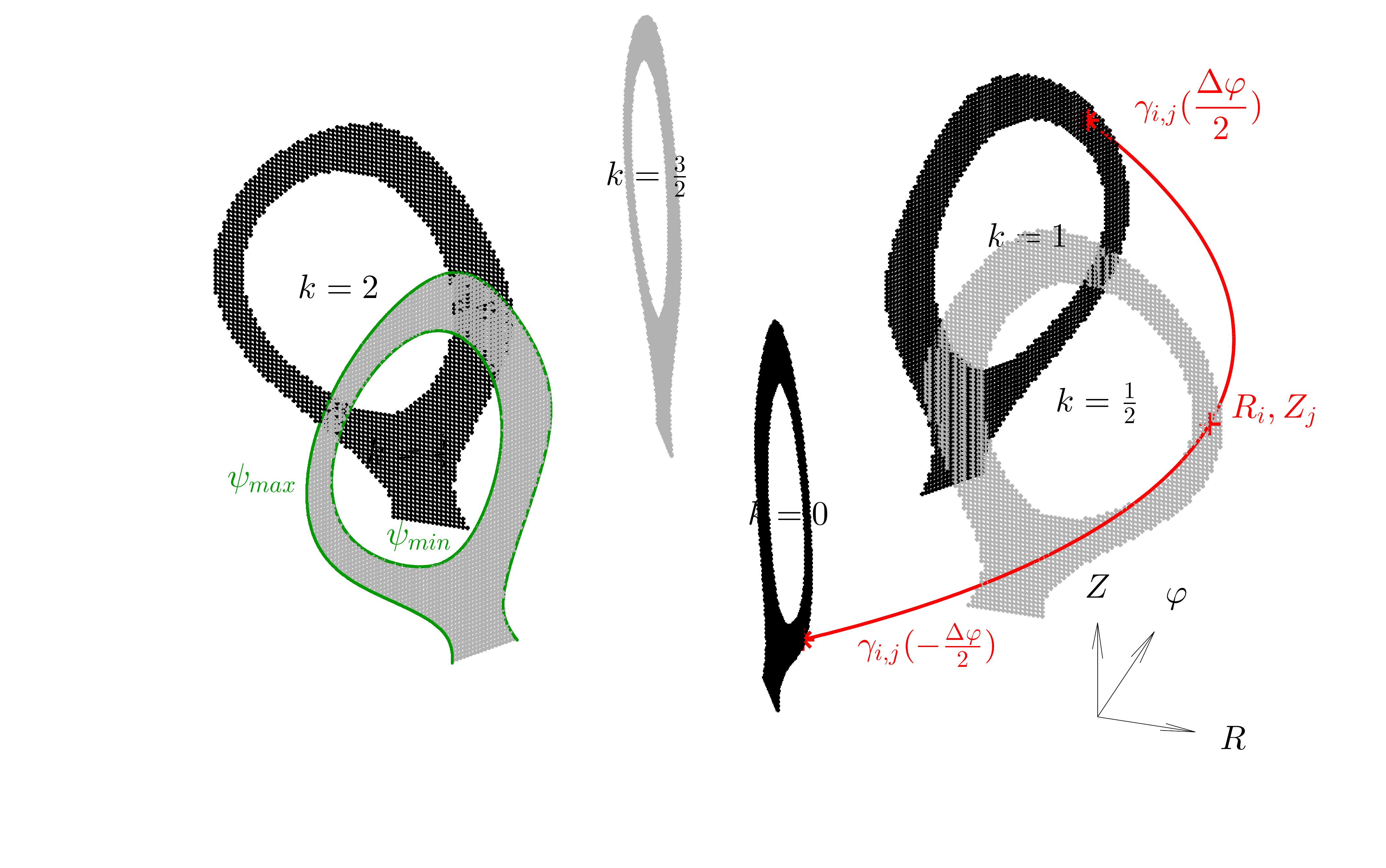}
\caption{Scheme for toroidally staggered FCI. A cylindric grid $(R_i,\varphi_k,Z_j)$ is used spanning the simulation domain by a set of Cartesian poloidal planes bounded by limiting flux surfaces $(\psi_{min},\psi_{max})$. In addition to the canonical grid (black, $k=0,1,2,\dots$)  a toroidally staggered dual grid (gray, $k=\frac{1}{2},\frac{3}{2},\frac{5}{2},\dots$) is introduced. The parallel gradient is discretised via field line tracing and interpolation and maps from the canonical grid to the staggered grid.}
\label{fig:fci_scheme_full}
\end{figure}

GRILLIX is based on the FCI approach \cite{hariri:fenicia13,stegmeir:cpc16,stegmeir:fciaddendum17} in a toroidally staggered framework which is described in detail in \cite{stegmeir:ppcf18} and is therefore here only reviewed very shortly. For tokamak geometries a cylindric grid $(R_i,\varphi_k,Z_j)$ is employed being Cartesian within poloidal planes. Based on the assumption of strong toroidal field ($B_{pol}/B_{tor}\ll 1$), the discretisation of perpendicular operators remains within poloidal planes for which second order finite difference methods are used. The Jacobi bracket is discretised according to the Arakawa scheme \cite{arakawa:scheme97} and the discretisation of the non-linear polarization term in eq.~(\ref{eq:vorticity}) is described in \cite{ross:nonbsq17}. The discrete parallel gradient is computed at toroidally staggered positions $\varphi_{k+\frac{1}{2}}$ according to a finite difference along magnetic field lines (see fig.~\ref{fig:fci_scheme_full}):
\begin{align*}
\nabla_\parallel f_{i,k+\frac{1}{2},j}:=\frac{f_{k+1}\left(\boldsymbol{\gamma}_{i,j}\left(\frac{\Delta\varphi}{2}\right)\right)-f_k\left(\boldsymbol{\gamma}_{i,j}\left(-\frac{\Delta\varphi}{2}\right)\right)}{s_{i,j}\left(\frac{\Delta\varphi}{2}\right)+s_{i,j}\left(-\frac{\Delta\varphi}{2}\right)},
\end{align*}
where $\boldsymbol{\gamma}_{i,j}(\varphi)$ is the poloidal projection of the characteristic along field line and $s_{i,j}(\varphi)$ the associated length along field line defined as the solution of the following ordinary differential equations which are solved via a Runge-Kutta integrator \cite{hairer:dop85393}:
\begin{align*}
\frac{d\boldsymbol{\gamma}_{i,j}}{d\varphi}=&\frac{1}{B^\varphi}\left(\begin{matrix}B^R \\B^Z\end{matrix}\right), & \text{with: }& \boldsymbol\gamma_{i,j}(0)=\left(\begin{matrix}R_i \\ Z_j\end{matrix}\right), \\
\frac{ds_{i,j}}{d\varphi}=&\frac{\left|B\right|}{B^\varphi}, &\text{with: }& s_{i,j}(0)=0,
\end{align*}
i.e.~corresponding map points are computed by tracing along magnetic field lines. The values on map points are obtained from a 3rd order bi-polynomial interpolation within the poloidal planes $\varphi_k$ and $\varphi_{k+1}$. We note that the magnetic field is assumed axisymmetric in GRILLIX, which is however not a general constraint for the FCI approach \cite{hill:fci17,shanahan:bsting18}. In the same spirit operators are established that map quantities between the grid and the staggered dual grid and vice versa. The parallel divergence operator $\nabla\cdot\left(\mathbf{b}f\right)$ is obtained in its discrete version via the support operator method \cite{shashkov:support95,shashkov:support96} as described in \cite{stegmeir:ppcf18}. The structure of the equations suggest that $n,\theta_n,\xi_e,T_e,\phi$ and $\Omega$ are co-located to the canonical grid whereas $u_\parallel,v_\parallel,j_\parallel,A_\parallel$ and $\psi_\parallel$ are co-located to the staggered grid. 

\subsection{Penalization for sheath boundary conditions}\label{subsec:penalization}
The boundaries at the sheath are in general neither conformal with the grid nor aligned with the exceptional parallel direction along the magnetic field line, which makes their treatment numerically difficult and cumbersome. In such situations penalization techniques have proven themselves also for plasma fluid codes \cite{isoardi:penalisation10,bufferand:edgesol2dpen13}. A combination of the FCI with penalization was firstly employed in GDB for limited plasmas \cite{zhu:gdb18,francisquez:phd18} and we implemented in GRILLIX a generalization allowing to deal also with diverted plasmas.

Equations (\ref{eq:continuity}-\ref{eq:etemperature}) are each modified according to:
\begin{align}
\frac{\partial}{\partial t}f=\left(1-\chi\right)F_f + \frac{\chi}{\epsilon}\left(f_P-f\right), 
\label{eq:penalisation_general}
\end{align}
where $f$ represents here the dynamical variables respectively and $F_f$ the corresponding terms according to the Braginskii model. $\chi$ is a characteristic function, which is $0$ in the physical domain and $1$ in the boundary region, where we choose in practice a smooth transition \cite{isoardi:penalisation10} across the boundary based on $\tanh$ functions (see fig.~\ref{fig:setup_chixi} left column for examples). $\epsilon\ll 1$ is the penalization parameter such that in the region where $\chi\approx0$ eq.~(\ref{eq:penalisation_general}) approximates the original physical equation, whereas in the region $\chi\approx 1$ the variable $f$ is strongly damped to a prescribed function $f_P$. Via suitable choice for $f_P$ different boundary conditions can be realized.

As an illustrative example we discuss here our implementation for general Neumann boundary conditions, i.e.:
\begin{align*}
\left.\nabla_\parallel f\right|_{sheath} = \alpha.
\end{align*}
Firstly, we define an additional function $\zeta$, which is $1$ in the penalization region where the magnetic field is pointed towards the target and $-1$, where the magnetic field is pointed away from the target with possibly, i.e.~in toroidal limiter geometry, a smooth transition between both regions (see fig.~\ref{fig:setup_chixi} right column for examples). Secondly, we denote for some grid point $f^\pm:=f_{k\pm1}\left(\boldsymbol{\gamma}_{i,j}(\pm\Delta\varphi)\right)$ the values on its map points which are again obtained via interpolation within adjacent planes and $s^\pm:=s_{i,j}(\pm\Delta\varphi)$ the associated lengths along field line. The penalization value is then prescribed as:
\begin{align}
f_P=\begin{cases}
\left|\zeta\right| \left(f^-+s^-\alpha\right) + (1-\left|\zeta\right|)\frac{f^++f^-}{2} & \text{for: } \zeta \ge 0,\\
\left|\zeta\right| \left(f^+-s^+\alpha\right) + (1-\left|\zeta\right|)\frac{f^++f^-}{2} & \text{for: } \zeta < 0.
\end{cases}
\label{eq:penalisation_neumann}
\end{align} 
The first terms set the actual boundary condition and use the values obtained from the field line map towards the interior domain. The second terms ensure for toroidal limiter geometries a continuous transition between both limiter sides.

A special treatment is needed for penalisation of the potentials. After having evolved the density, the electron temperature and the vorticity from time step $t$ to $t+1$ (see section \ref{subsec:time stepping}) the electrostatic potential $\phi^{t+1}$ is computed at time step $t+1$ according to:
\begin{align}
\frac{\chi}{\epsilon}\phi^{t+1}-\left(1-\chi\right)\nabla\cdot\left(\frac{n^{t+1}}{B^2}\nabla_\perp\phi^{t+1}\right) = \frac{\chi}{\epsilon}\left(\Lambda T_e^{t+1}\right)-\left(1-\chi\right)\Omega^{t+1},
\label{eq:potsolvepen}
\end{align}
yielding $\phi=\Lambda T_e$ in the penalisation region, where $\chi\approx 1$. The insulating sheath boundary condition implies that the generalised electromagnetic potential has to be penalised to $\psi_\parallel=0$, which is realised by adding the corresponding penalisation term to Ohm's law (\ref{eq:ohm}). After having evolved the generalised electromagnetic potential in time to $\psi_\parallel^{t+1}$ the electromagnetic potential $A_\parallel^{t+1}$ is obtained according to:
\begin{align}
\beta_0A_\parallel^{t+1}-\frac{\mu}{n^{t+1}}\nabla_\perp^2 A_\parallel^{t+1}=\psi_\parallel^{t+1},
\label{eq:aparsolve}
\end{align}
from which the parallel current is computed according to $j_\parallel^{t+1}=-\nabla_\perp^2A_\parallel^{t+1}$. Equations (\ref{eq:potsolvepen}) and (\ref{eq:aparsolve}) are Helmholtz equations for $\phi^{t+1}$ and $A_\parallel^{t+1}$ that are solved in GRILLIX via a multigrid solver (see section \ref{subsec:elliptic solver}). Finally, we note that we also tried slightly different methods for penalisation of the potentials \cite{francisquez:phd18}, and the results seemed not to depend strongly on the details of the technique employed.

\subsection{Time stepping}\label{subsec:time stepping}
The equations are advanced in time with the 3rd order Karniadakis scheme \cite{karniadakis:bdf91}. Only the penalization term that is directly proportional to the quantity itself is treated fully implicit in time, i.e.~the equations written in the form of eq.~(\ref{eq:penalisation_general}) are discretised in time according to:
\begin{align*}
f^{t+1}\left(11+6\Delta t\frac{\chi}{\epsilon}\right) = \sum\limits_{i=0\dots2}a_if^{t-i}+b_i\Delta t\left[\left(1-\frac{\chi}{\epsilon}\right)F^{t-i}(f^{t-i})+\frac{\chi}{\epsilon}f_P^{t-i}\right],
\end{align*}
with $a_0=18,\,a_1=-9,\,a_2=2$ and $b_0=3,\,b_1=-3,\,b_2=1$. The solution for $f^{t+1}$ is trivial as the implicit penalization term on the left hand side is diagonal.

\subsection{Elliptic solver}\label{subsec:elliptic solver}
In order to compute the electrostatic potential $\phi$ from eq.~(\ref{eq:potsolvepen}) and the parallel electromagnetic potential $A_{\parallel}$ from eqs.~(\ref{eq:aparsolve}) two Helmholtz type equations have to be solved in each time step within each poloidal plane:
\begin{align}
c_1 f-c_2\nabla\cdot\left(c_3\nabla_\perp f \right)=b,
\label{eq:helmholtz}
\end{align}
with given right hand side $b$, and coefficients $c_1,\,c_2$ and $c_3$. In the global model the coefficients have in general a spatio-temporal dependency, where direct solvers become very inefficient as a costly matrix LU-decomposition would have to be performed in each time step. An efficient solution technique for eq.~(\ref{eq:helmholtz}) is provided by geometric multigrid methods \cite{hackbusch:mgrid85}, which is implemented in GRILLIX based on a damped Jacobi smoother with trivial restriction and bilinear prolongation.

\subsection{Verification}
\begin{figure}
\centering
\includegraphics[trim=0 0 0 0,clip,width=0.75\linewidth]{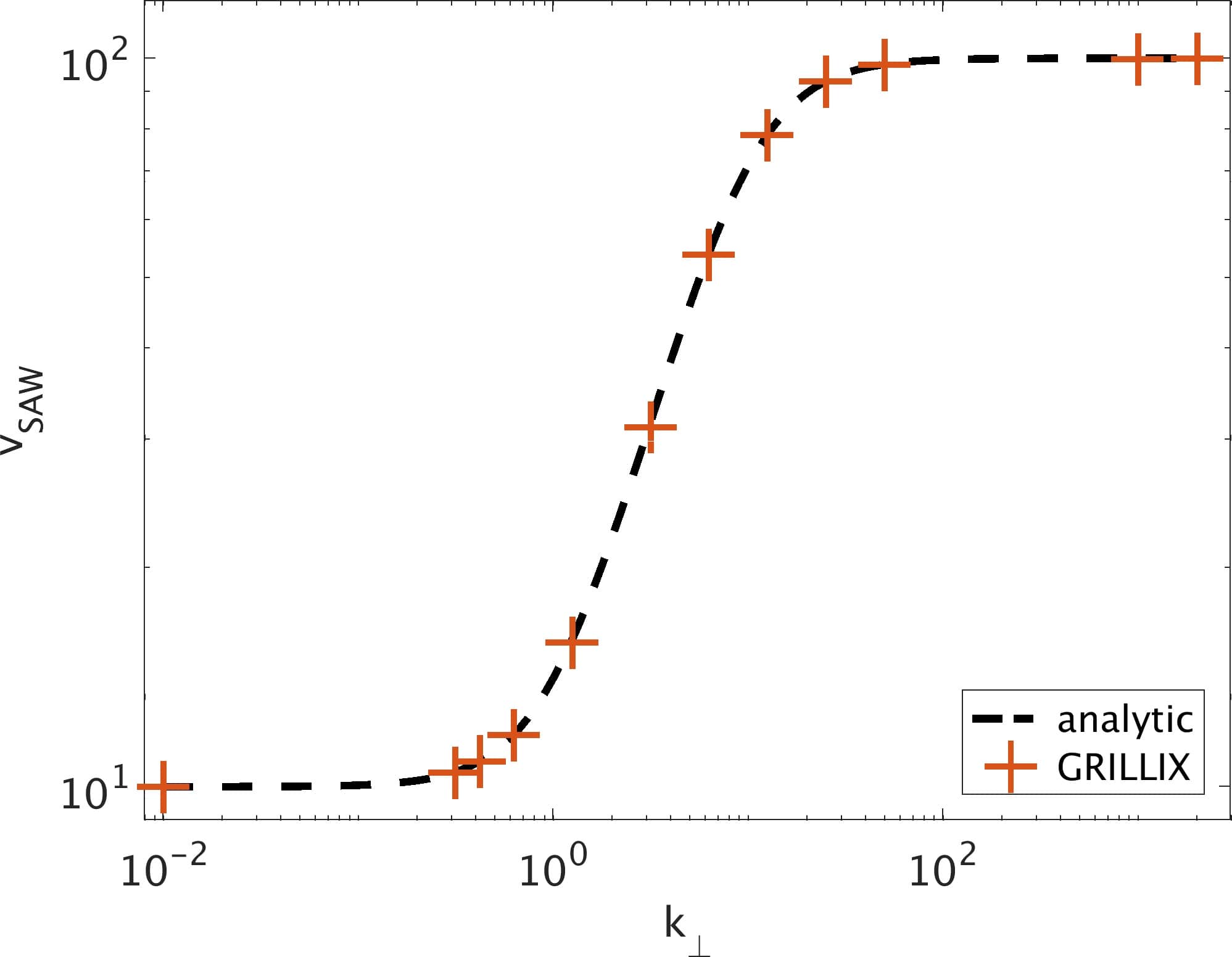}
\caption{Velocity of shear Alfv\'en wave obtained with GRILLIX against analytic prediction for fixed $\beta_0=1\cdot10^{-2}$ and $\mu=1\cdot10^{-4}$ in dependence of $k_\perp$}
\label{fig:alfven_kperp}
\end{figure}

One of the main new features in GRILLIX are electromagnetic and electron inertial effects in Ohm's law, which gives rise to shear Alfv\'en dynamics. The core model for the shear Alfv\'en wave is obtained by linearizing equations (\ref{eq:continuity}-\ref{eq:faraday}) in the isothermal limit $(T_e=1)$, neglect curvature, parallel ion velocity $(u_\parallel=0)$ and parallel resistivity $\eta_{\parallel0}=0$. In this limit a wave equation is obtained \cite{scott:habil00} for the `non-adiabaticity` $(\tilde{n}-\tilde{\phi})$, where the tilde denotes a fluctuating quantity.
\begin{align}
\frac{\partial^2}{\partial t^2}\left(\tilde{n}-\tilde{\phi}\right)=v_{SAW}^2\nabla_\parallel^2\left(\tilde{n}-\tilde{\phi}\right),
\end{align}
with v$_{SAW}^2=(1+k_\perp^2)/(\beta_0+\mu k_\perp^2)$ the phase velocity of the shear Alfv\'en wave, where $k_\perp$ is the perpendicular mode number. In the limit $k_\perp\ll 1$ the wave propagates at the Alfv\'en speed $v_A=\beta_0^{-1/2}$ and in the limit $k_\perp\gg 1$ at the electron thermal speed $v_{Te}=\mu^{-1/2}$. In order to verify the implementation of the electromagnetic and electron inertia effects we perform simulations with GRILLIX in a 3D periodic slab $(\mathcal{C}=0)$ without the parallel momentum equation (\ref{eq:parmomentum}) and electron thermal equation (\ref{eq:etemperature}), but set $T_e=1$ and $u_\parallel=0$. Otherwise we run the global version of the code but initialize the density with constant background plus a fluctuation of small amplitude $\frac{\tilde{n}}{n_{bck}}=0.1$ being a mode structure in the perpendicular plane and a Gaussian along the parallel direction. The phase velocity of the divergent wave along the magnetic field is measured and compared to the analytic prediction for $v_{SAW}$. The result in fig.~\ref{fig:alfven_kperp} shows an excellent agreement between GRILLIX simulations and the analytic prediction.

The presence of shear Alfv\'en dynamics is also numerically beneficial, as it limits the parallel electron motion resulting in a Courant-Friedrichs-Lewy number (CFL) based on the Alf\'en speed. In the electrostatic case $(\beta_0=0,\,\mu=0)$ parallel electron motion would only be hindered by parallel resistivity. A rough guess on the time step limitation in the electrostatic case can be obtained by considering the linearised vorticity equation (\ref{eq:vorticity}) and Ohm's law (\ref{eq:ohm}) neglecting electron and density variations, i.e.~$n=1,\,T_e=1$. For a single perpendicular mode $\nabla_\perp^2\rightarrow -k_\perp^2$ a parallel diffusion equation for the electrostatic potential is finally obtained:
\begin{align*}
\frac{\partial}{\partial t}\phi=\frac{1}{k_\perp^2\eta_\parallel}\nabla\cdot\left(\mathbf{b}\nabla_\parallel\phi\right).
\end{align*}
Treating this problem explicit in time would result in a time step limitation of $\Delta t\lesssim\Delta s^2 k_\perp^2\eta_\parallel$, where $\Delta s$ is the parallel grid distance. As this time step limitation is severe for large modes at low resistivity, it would make an implicit treatment practically inevitable \cite{tamain:tokam3x16,scott:dwtimplicit88}.

\begin{figure}
\includegraphics[trim=0 0 0 0,clip,width=1.0\linewidth]{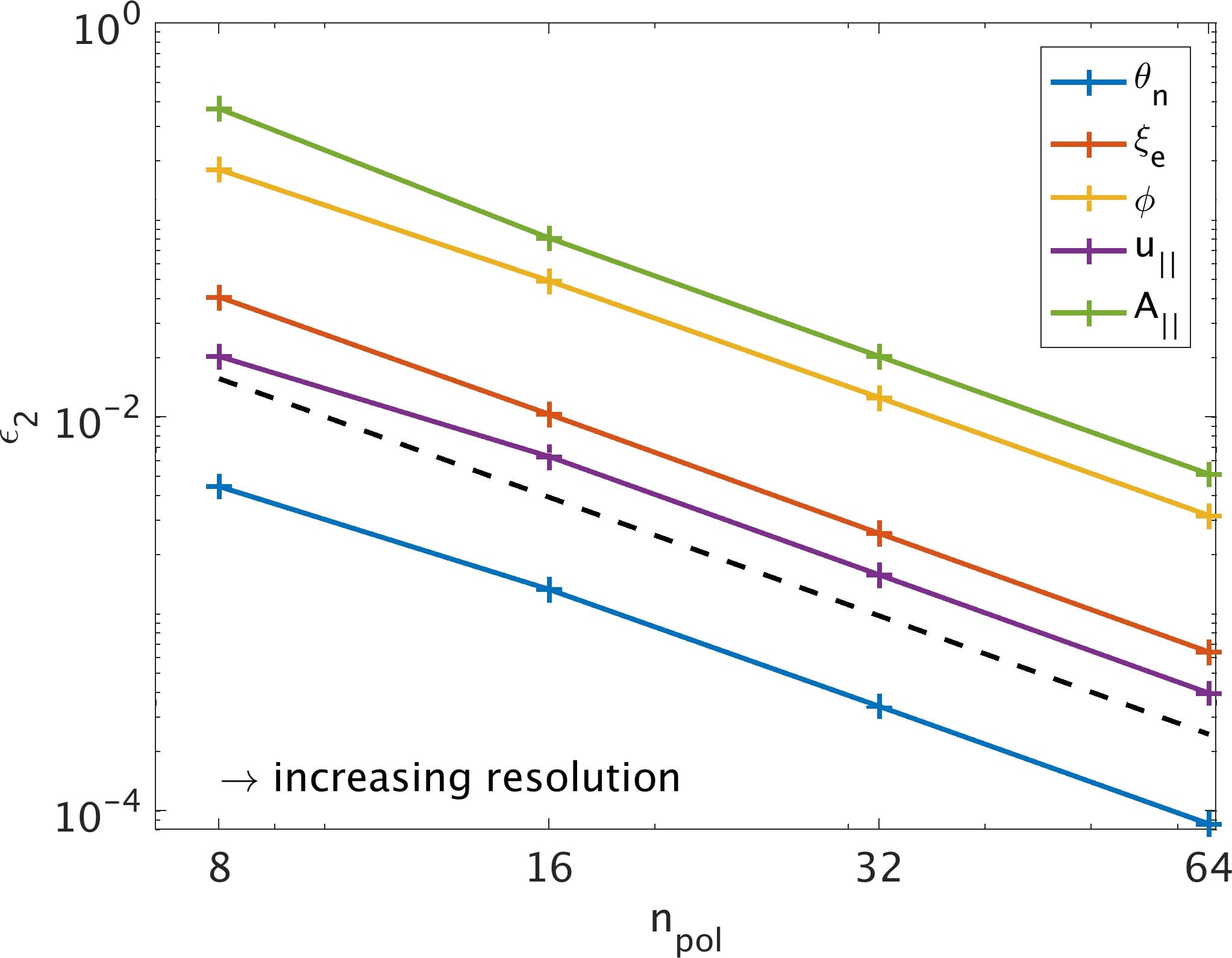}
\caption{Numerical error of MMS verification procedure for different dynamical fields evolved in GRILLIX. Error is measured in L2-norm, i.e.~$\varepsilon_2=\left|u_{num}-u_{mms}\right|_2/\left|u_{mms}\right|_2$ at $t=0.2$. Black dashed lines indicates second order convergence for reference. Resolution is subsequently doubled in all directions starting at the coarsest level with $n_{pol}=8$ poloidal planes, $h=3\cdot10^{-3}$ and $\Delta t=1\cdot10^{-4}$.}
\label{fig:mms_circ}
\end{figure}

A verification of the full system via the Method of Manufactured Solution (MMS)  \cite{salari:mms00} was performed in circular geometry with purely closed flux surfaces. The same setup and procedure as described in \cite{stegmeir:ppcf18} for the previous simplified set of equations is applied here to the global model. The analytic MMS functions are prescribed for each dynamical field as a product of radial ($k_\rho$), poloidal ($k_\theta$ with phase shift $\delta\theta$), toroidal ($k_\varphi$ with phase shift $\delta\varphi$) and temporal ($\omega$) modes (see table \ref{table:mms_parameters}). The analytic MMS functions are quite general for GRILLIX, as its numerical approach is independent of flux surfaces. The numerical error of the MMS analysis for all dynamical fields in dependence of resolution is shown in fig.~\ref{fig:mms_circ} and follows a second order convergence, which is a good indication for correct implementation of the equations in GRILLIX.

\begin{table}
\centering
\begin{tabular}{l|cccc}
              & $k_\rho$ & $(k_\theta,\delta\theta)$ & $(k_\varphi,\delta\varphi)$ & $\omega$ \\
\hline
$\theta_n$    & $ 1 $  &  $ (1,\,0)   $  & $ (1,\,0)   $ & $ 100 $  \\
$\xi_e$       & $ 1 $  &  $ (1,\,1.5) $  & $ (1,\,0.5) $ & $ 73  $  \\
$\phi$        & $ 2 $  &  $ (2,\,0)   $  & $ (1,\,0)   $ & $ 80  $  \\
$u_\parallel$ & $ 3 $  &  $ (1,\,0)   $  & $ (1,\,0)   $ & $ 65  $  \\
$A_\parallel$ & $ 2 $  &  $ (1,\,0)   $  & $ (1,\,0)   $ & $ 88  $    
\end{tabular}
\caption{Parameters for analytic MMS functions used as inputs for different dynamical fields.}
\label{table:mms_parameters}
\end{table}

Finally, we want to note that also a validation in slab geometry based on experiments in the Large Plasma Device (LAPD) was performed for which results can be found in \cite{ross:phd18,ross:bossinesq18}.

\section{Simulation results}\label{sec:Simulation_results}

\subsection{Setup}
The parameters for our simulations are motivated from experiments with deuterium plasma in the COMPASS tokamak \cite{panek:compass15}, where we normalise density and temperature to upstream separatrix values:
\begin{align*}
R_0 =& 56\,\text{cm}, & a_{min} =& 20\,\text{cm}, & B_0 =& 1.2\,\text{T} ,\\ n_0 =& 1\cdot10^{13}\,\text{cm}^{-3}, & T_0 =& 30\,\text{eV},
\end{align*}
with $a_{min}$ the minor radius. This corresponds to the following dimensionless input parameters for GRILLIX:
\begin{align*}
\beta_0 =& 1.68\cdot10^{-4}, & \delta =&849, & \mu =&2.72\cdot10^{-4}, \\ \chi_{\parallel0} =& 340, & \eta_{\parallel0} =& 4.72\cdot10^{-3}. 
\end{align*}
The strongest time step limitation stems from the parallel electron heat conduction, and by linearizing this term we may anticipate a rough scaling for the time step of $\Delta t \lesssim \frac{n}{\chi_{\parallel0}T_e^{5/2}}\Delta\varphi^2$ with $\Delta\varphi$ the toroidal grid distance between adjacent planes. In order to perform simulations at a larger time step we employ in our simulations presented here mostly a reduced heat conductivity of $\chi_{\parallel0}=20$, but still retain the parametric dependency on temperature ($\propto T_e^{5/2}$). Correspondingly, we also use a reduced value for the effective sheath transmission factor of $\gamma_e=0.15$. In section \ref{subsec:convergenc_and_heatconductivity} we investigate the effect of this by comparing to a run with more realistic heat conductivity. Development towards relaxation of the time step limitation  by an implicit treatment of parallel electron heat conduction is targeted for future work.

\begin{figure}
\includegraphics[trim=0 0 0 0,clip,width=1.0\linewidth]{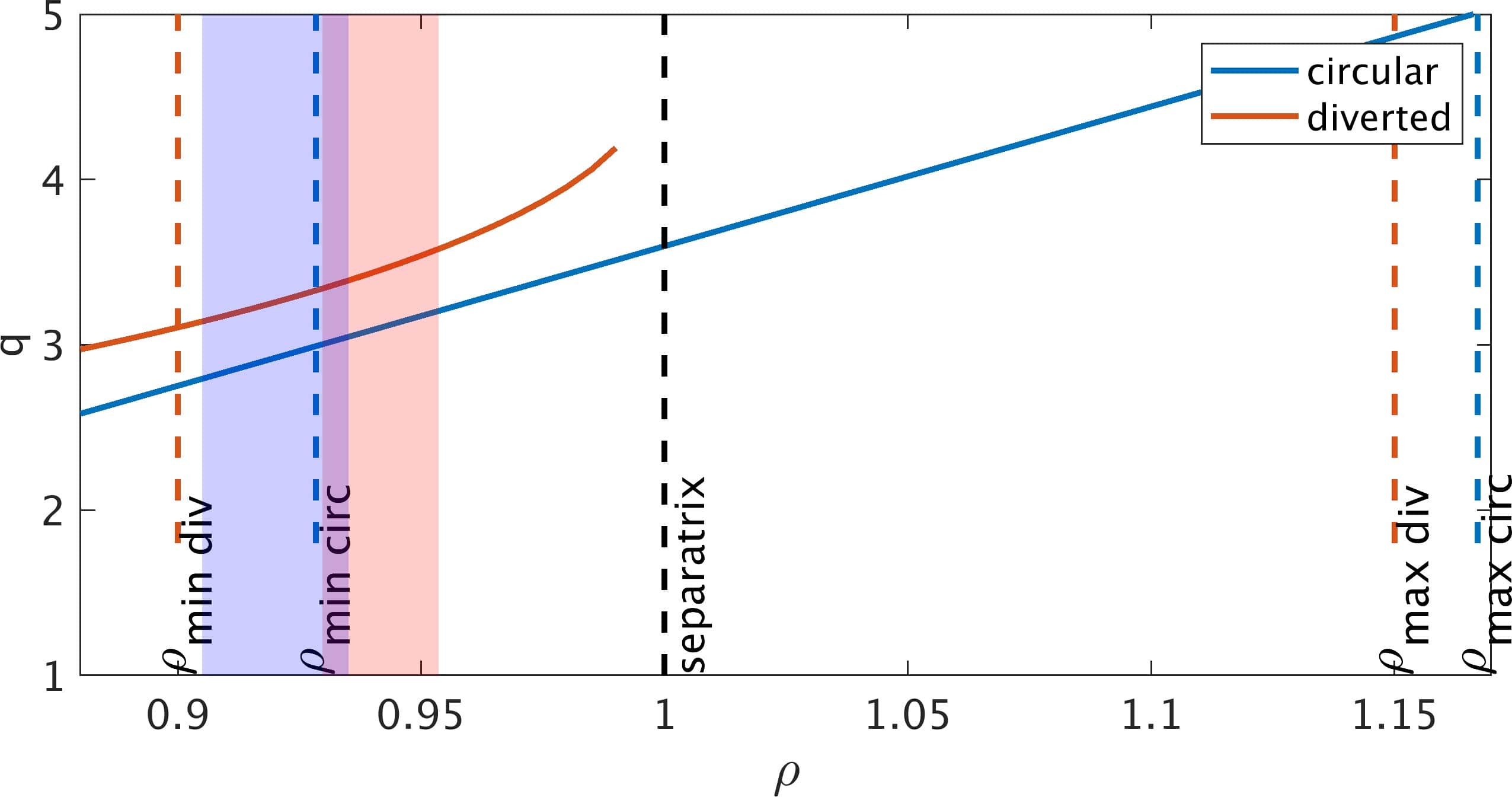}
\caption{Safety factor $q$ as function of normalised radial coordinate $\rho$ for circular and diverted geometry. The limiting flux surfaces are indicated with dashed blue respectively red lines and the source regions with coloured shaded areas.}
\label{fig:setup_radial}
\end{figure}

\begin{figure}
\centering
\includegraphics[trim=0 0 0 0,clip,width=0.9\linewidth]{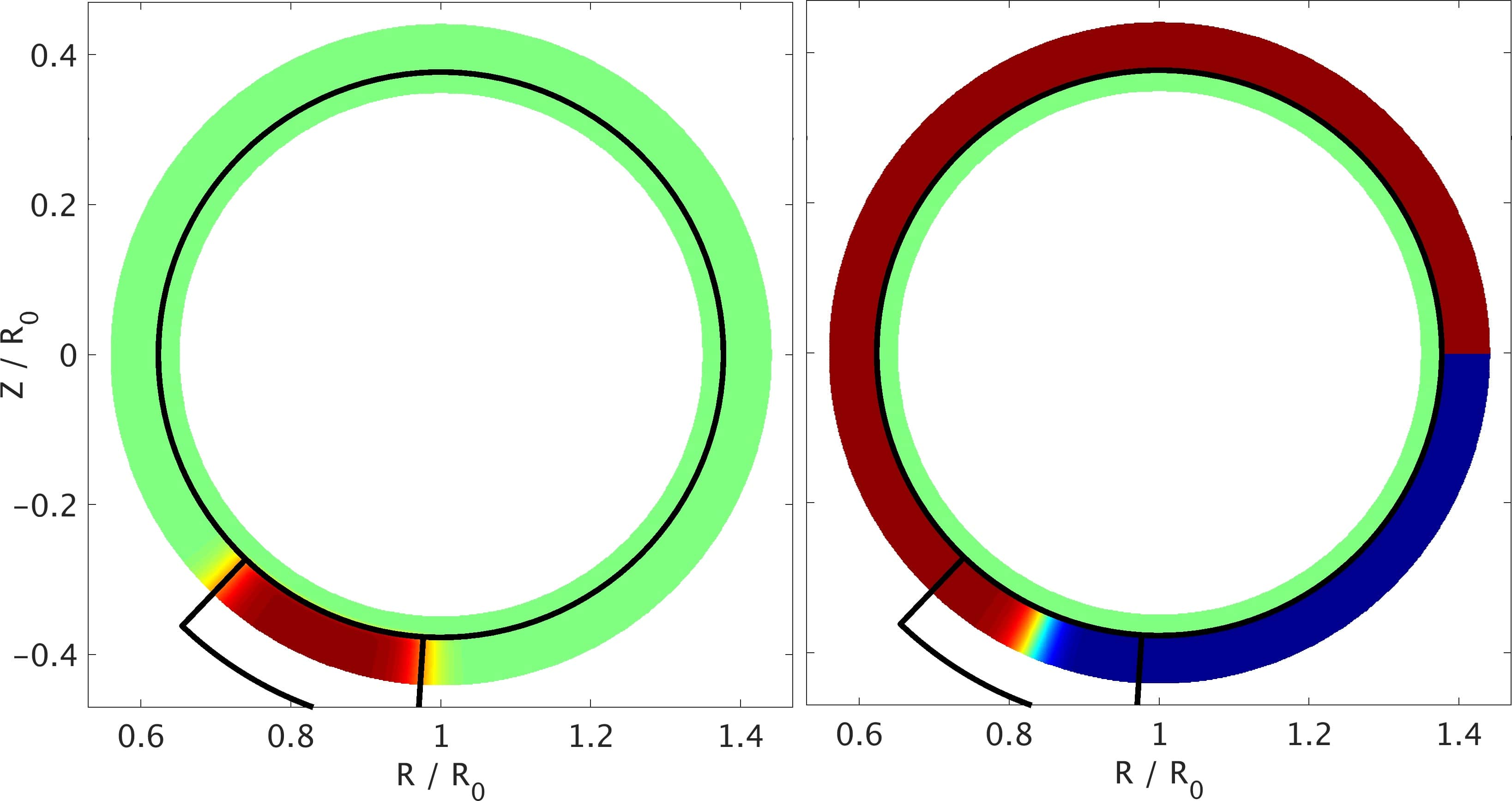}
\includegraphics[trim=0 0 0 0,clip,width=0.75\linewidth]{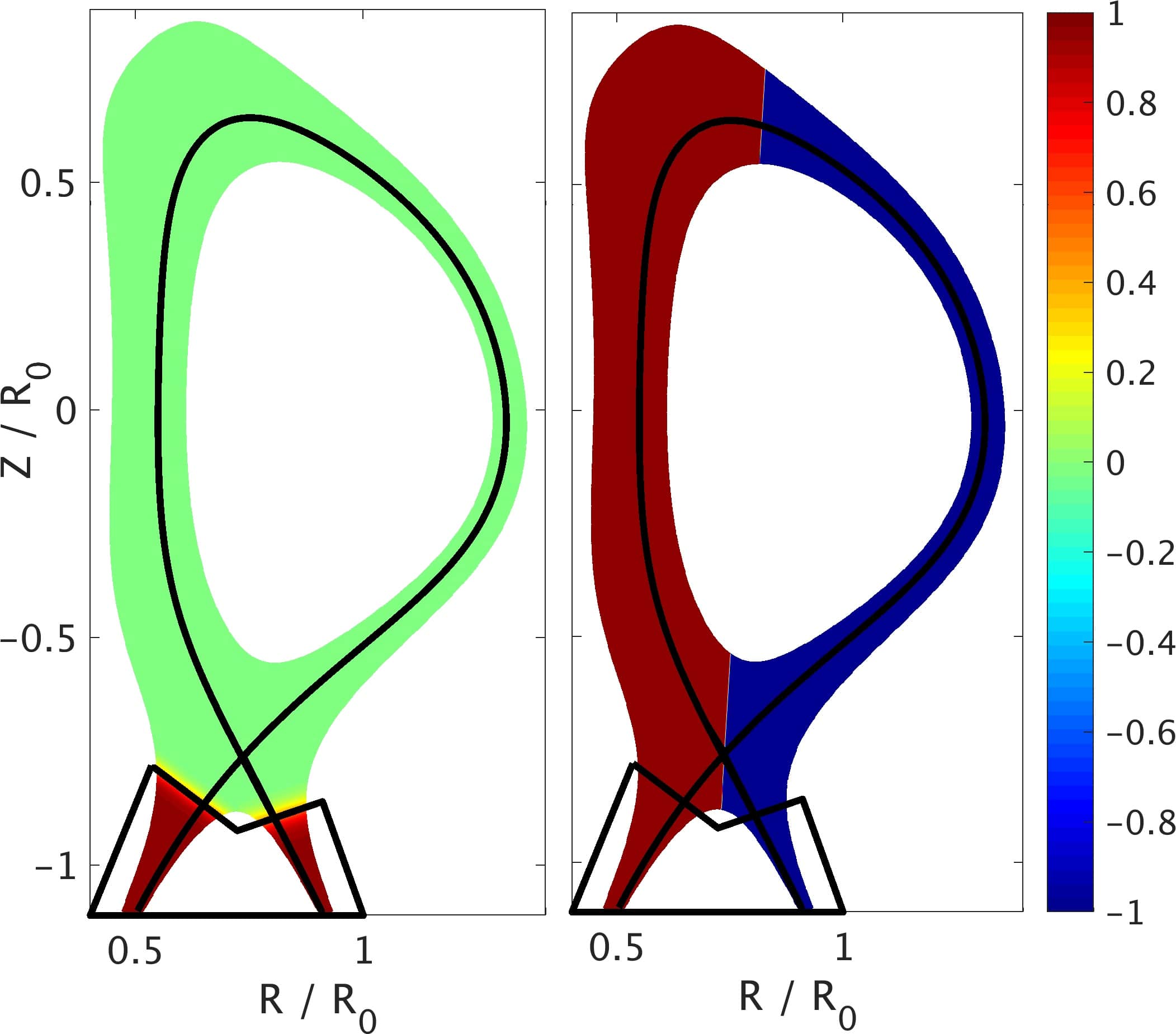}
\caption{Penalization functions $\chi$ (left) and $\zeta$ (right) used in circular (top) and diverted (bottom) geometry. Black lines indicate separatrix and limiter respectively divertor target plates.}
\label{fig:setup_chixi}
\end{figure}

We performed simulations in circular geometry with toroidal limiter and diverted geometry at otherwise comparable parameters. The background magnetic field for the diverted geometry is given in terms of an analytic flux function $\psi(R,Z)$ from the class of solutions described in \cite{cerfon:equilibrium10} with  parameters chosen as described in \cite{held:cpc16}. We define as normalised radial coordinate $\rho=\sqrt{\frac{\psi-\psi_0}{\psi_X-\psi_0}}$, where $\psi_0,\,\psi_x$ is the poloidal magnetic flux at magnetic axis respectively at separatrix. For circular geometry we define equivalently the normalised radial coordinate as $\rho:=\frac{\sqrt{(R-R_0)^2 + Z^2}}{a_{min}}$ and the magnetic field is given in terms of a prescribed safety factor profile $q(\rho)$. The setup for the simulations in terms of radial view is shown in fig.~\ref{fig:setup_radial} and the characteristic functions $\chi$ prescribing the location of the sheath via penalisation is illustrated for both geometries in fig.~\ref{fig:setup_chixi}. 

The simulations are driven via a particle source of the form
\begin{align*}
S_{n}=c_n\exp\left(-\frac{(\rho-\rho_{src})^2}{w_{src}^2}\right)\left(\left\langle n\right\rangle_{\rho}-n_{target}\right)
\end{align*}
and an energy source $S_T$ of the same form. The sources are located near the inner boundary at $\rho_{src}$ and drive the zonal averaged density $\left\langle n\right\rangle_{\rho}$ and temperature $\left\langle T_e\right\rangle_{\rho}$ towards prescribed values $n_{target}$ and $T_{e,target}$ within a narrow region $(w_{src})$. The parameters $c_n,\,c_t$ control the rate of the sources. Outside the source region the profiles relax freely and develop self-consistently. We preferred this form for the sources to a purely flux driven source as it allows an effective control over keeping simulations within a desired parameter regime. Eventually, it is comparable to other approaches, who source their simulations via penalising the profiles near the core \cite{francisquez:phd18} or use a feedback control loop \cite{dudson:hermes17}.

The main simulations analysed in section \ref{subsec:circturb} to \ref{subsec:bsqcomparison} were run with $32$ poloidal planes, perpendicular resolution of $h= 1\, [\rho_{s0}]$ corresponding to $0.066\text{ cm}$ and a time step of $\Delta t=5\cdot10^{-5}\,[R_0/c_{s0}]$. The total number of grid pints were $\approx5.0\cdot10^6$ grid points for the circular case and $\approx14\cdot10^6$ for the diverted case. A sixth order hyperviscosity $(\nu_{\perp,f}\nabla_\perp^6)$ is applied in the perpendicular direction and regular diffusion $(\mu_{\parallel,f}\nabla\cdot\left(\mathbf{b}\nabla_\parallel\right))$ in the parallel direction, where the coefficients were chosen as $\nu_{\perp,f}=10$ and $\mu_{\parallel,f}=0.025$ cutting off turbulent spectra by smoothing structures on the grid scale. The independence of the results from these numerical parameters was checked at the circular case (see section \ref{subsec:convergenc_and_heatconductivity}). The radial extent of the simulations in circular geometry is $\sim 1.5\text{ cm (edge)}\, +\, 3.5\text{ cm (SOL)}$ and in diverted geometry $\sim 1.5\text{ cm (edge)}\, +\, 2.3\text{ cm (SOL)}$ at outboard mid plane respectively  $\sim 3.2\text{ cm (edge)}\, +\, 5.5\text{ cm (SOL)}$ at inboard mid plane.

The simulations were initialized with uniform background in density and temperature $(0.2)$ plus small random noise $(0.01)$. Particles and energy are injected via the sources, and the simulations enter saturated phase, which is independent of the initial state (for the circular case around $t\approx 30$) from where data is collected for performing statistical analysis. The overall simulation time  for the circular case was up to $t=77$, corresponding to $\approx 1\text{ms}$. The simulations were carried out on the Marconi-A2 (KNL) partition on 16 nodes (2 MPI processes times 34 cores per node). Within 24 hours GRILLIX ran a normalised time interval of $\approx10\,[R_0/c_{s0}]$ for the circular case and $\approx4\,[R_0/c_{s0}]$ for the diverted case.

\subsection{Circular geometry}\label{subsec:circturb}
\begin{figure}
\includegraphics[trim=0 0 0 0,clip,width=1.0\linewidth]{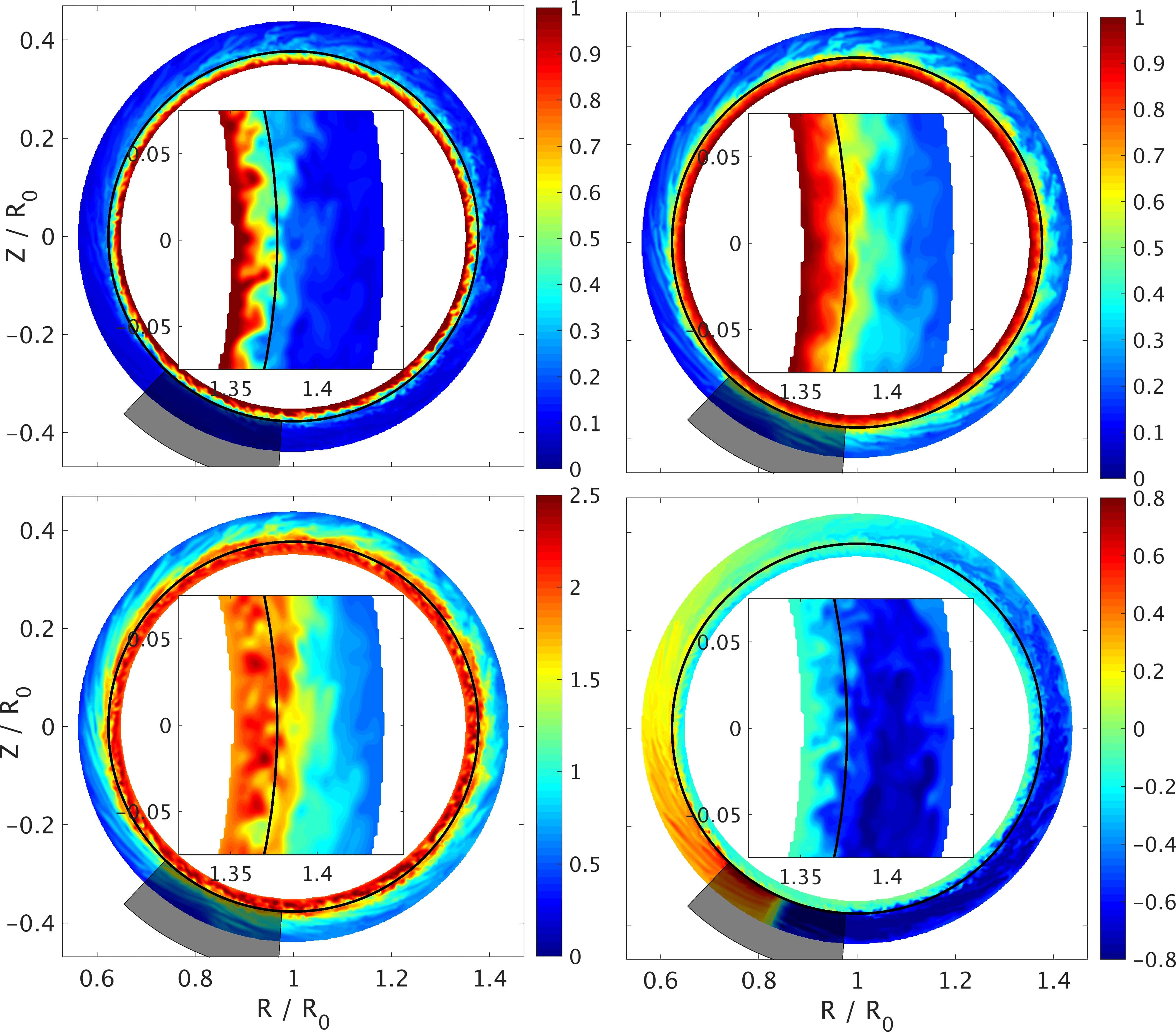}
\caption{Snapshots of density (top left), electron temperature (top right), electrostatic potential (bottom left) and parallel ion velocity (bottom right) in circular geometry. Insets show outboard mid plane region enlarged. Solid black line indicates last closed flux surface and gray shaded area penalization region due to toroidal limiter.}
\label{fig:csnaps}
\end{figure}

Snapshots of density, electron temperature, electrostatic potential and parallel velocity for the circular geometry with toroidal limiter are shown in fig.~\ref{fig:csnaps}. There is a clear difference between the closed field line region and the SOL, which is dominated by the Bohm boundary condition for the parallel velocity. Blob-like structures in the density can be observed around the last closed flux surface. 

\begin{figure}
\includegraphics[trim=0 0 0 0,clip,width=1.0\linewidth]{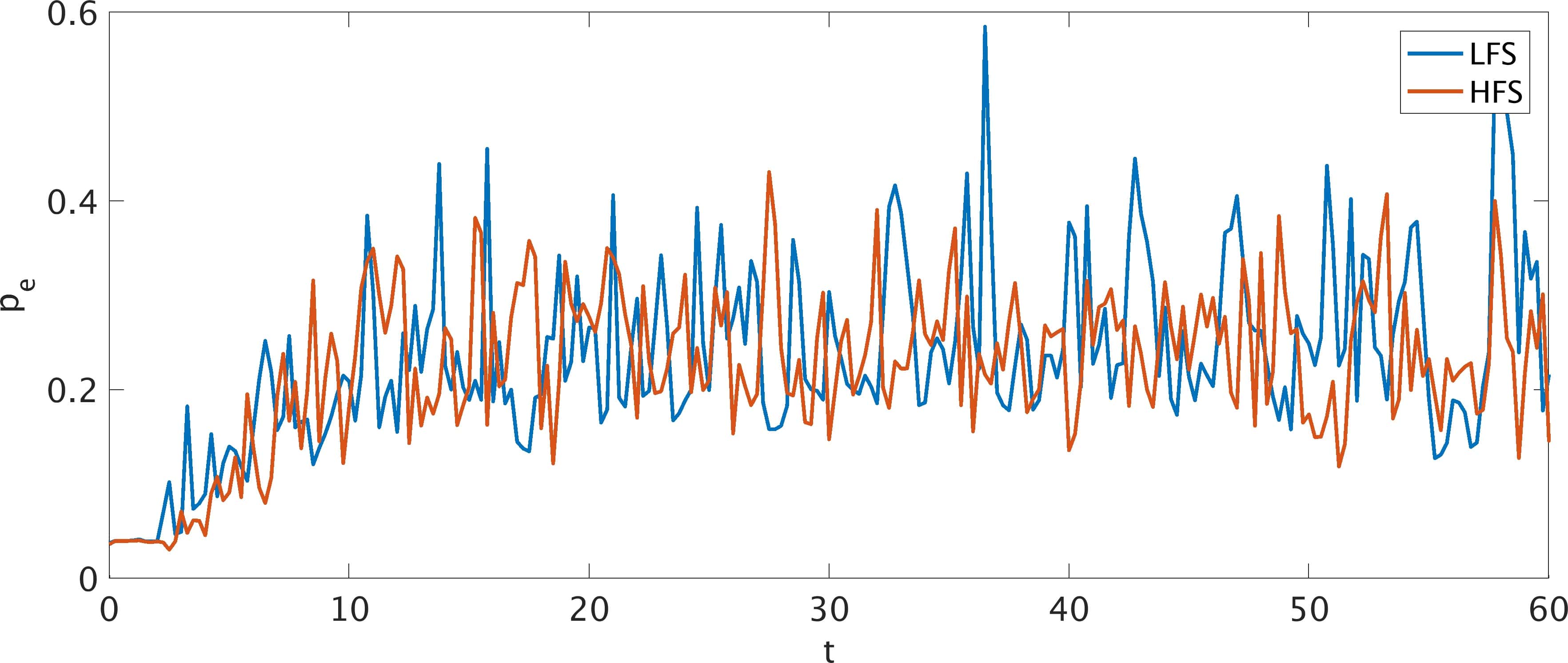}
\caption{Time traces of pressure on the last closed flux surface $(\rho=1.0)$ at low (LFS) and high (HFS) field side mid plane positions for circular limited geometry.} 
\label{fig:ctraces}
\end{figure}

\begin{figure}
\includegraphics[trim=0 0 0 0,clip,width=1.0\linewidth]{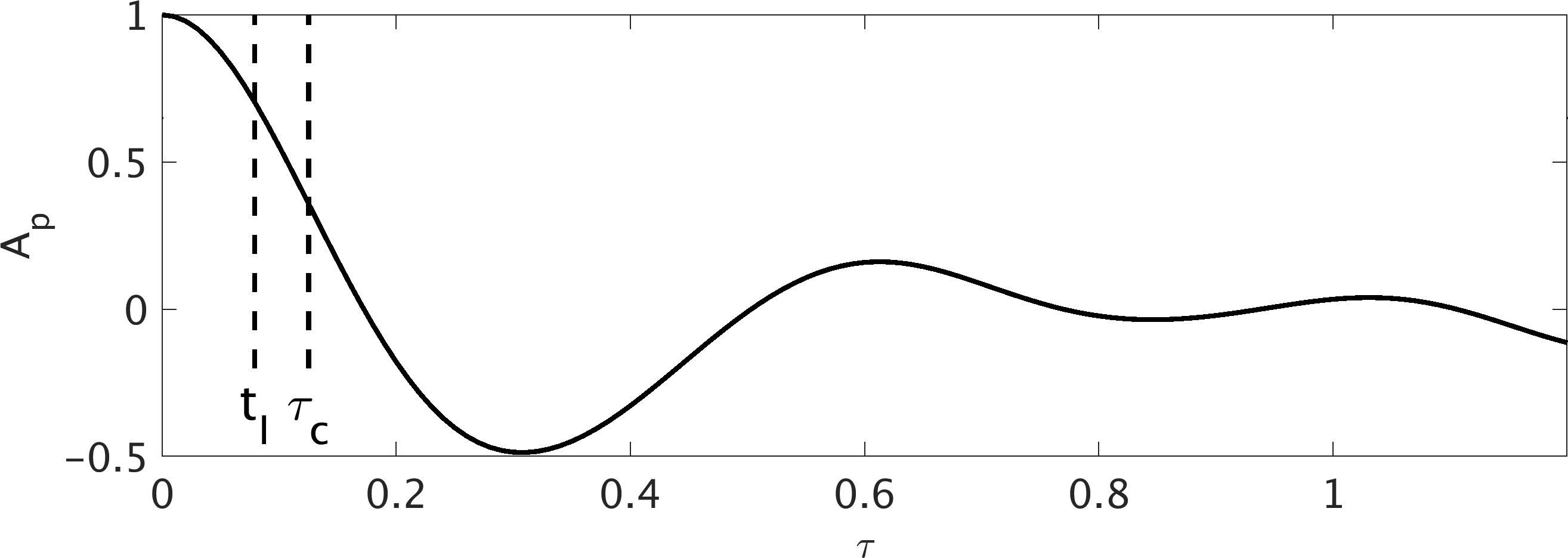}
\caption{Autocorrelation function of pressure signal taken at low field side in closed field line region where pressure gradient is strongest at (at $(R-R_{LCFS})/\rho_{s0}=-11$, see fig.~\ref{fig:profiles_circular}). Right dashed vertical lines indicates autocorrelation time $\tau_c$, where autocorrelation function drops to $1/e$ of its peak value, and left vertical line indicates characteristic time-scale related with growth rate of interchange instability $t_I=\sqrt{L_p/(2R_0)}$.}
\label{fig:cacorr}
\end{figure}

\begin{figure}
\includegraphics[trim=0 0 0 0,clip,width=1.0\linewidth]{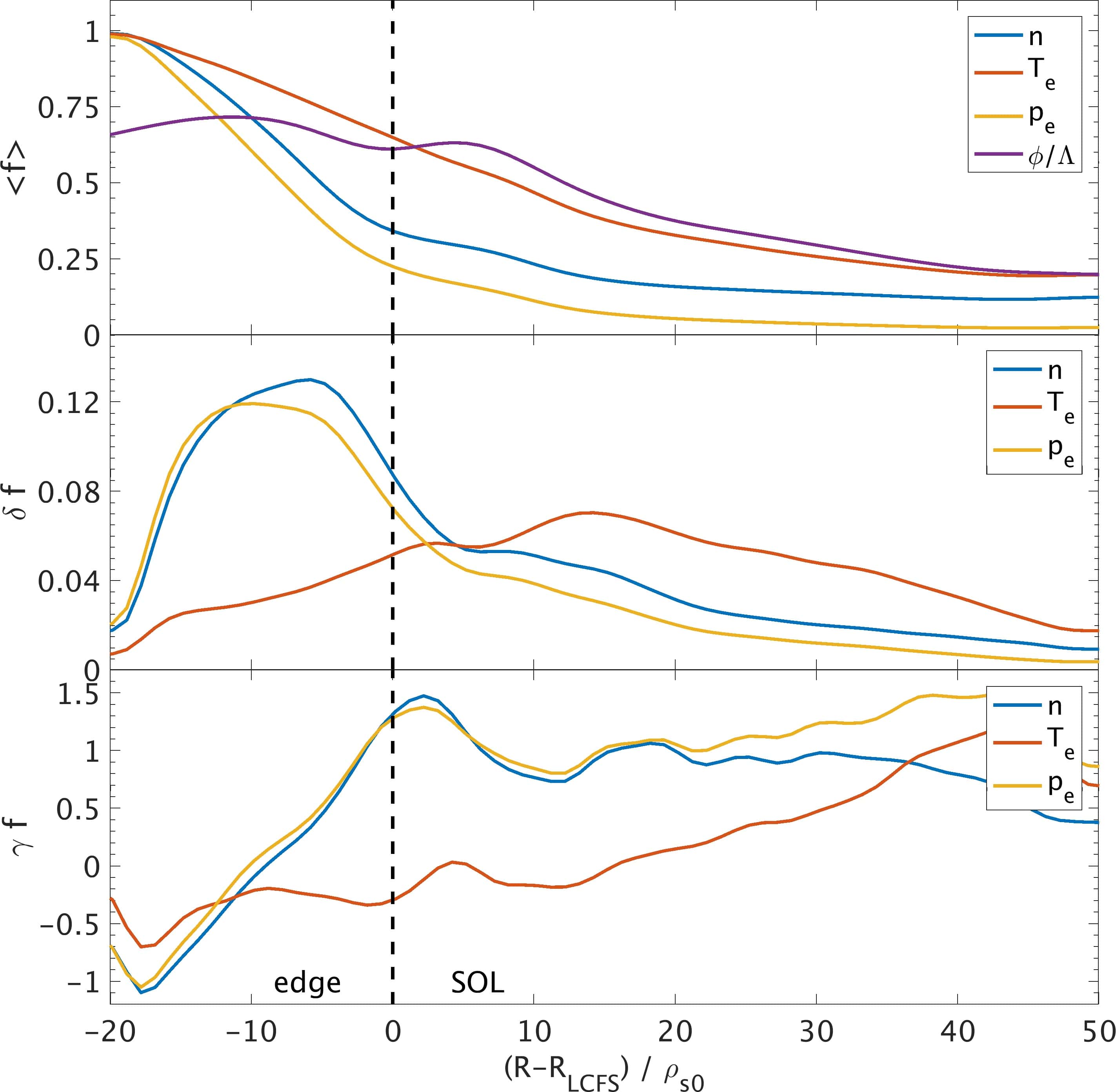}
\caption{Radial cut at outboard mid plane showing profiles (top), fluctuation levels (center) and skewnesses (bottom) for circular geometry. Dashed black line indicates last closed flux surface.}
\label{fig:profiles_circular}
\end{figure}

Time traces of pressure during the saturated state at low (LFS) and high (HFS) field side mid plane are shown in fig.~\ref{fig:ctraces}. Especially the LFS-signal is highly intermittent with fluctuations of up to $200\%$, implying that a global model, which does not rely on a splitting into fluctuations and background, is indeed important. The fluctuation level on the HFS is slightly lower which results from the ballooning character of the turbulence where curvature acts stabilizing at HFS and destabilizing at LFS. It has been found that turbulence in the SOL of limited plasmas is driven by resistive ballooning modes \cite{halpern:solballooning13,zhu:gdb17} with linear growth rates somewhat smaller than the interchange growth rate $\gamma_I=\sqrt{2R_0/L_p}$, where $L_p$ is the background pressure gradient length. Via the autocorrelation we may gain some insight into the characteristic time scales of the turbulent dynamics, which is computed discretely and in normalised form as:
\begin{align*}
A_f(\tau_i):=\frac{\sum\limits_{n}f(t_n)f(t_{n-i})}{\sum\limits_{n}f(t_n)^2},
\end{align*}
with $f(t_n)$ the signal at discrete time point $t_n$. The autocorrelation function for the pressure at LFS in the region of strongest pressure gradient is shown in fig.~\ref{fig:cacorr}, and a correlation time of $\tau_c\approx0.125$ is obtained. The characteristic turbulent time scales are therefore slightly larger than the interchange time scale $t_I=\gamma_I^{-1}\approx 0.08$, where the pressure gradient length in the edge has been estimated from the self-consistently obtained profiles (see fig.~\ref{fig:profiles_circular}) as $L_p/R_0\approx 0.012$. The turbulence is therefore compatible to be driven by resistive ballooning modes.

Furthermore, we analyse the simulation by computing profiles $\left\langle f\right\rangle$, fluctuation levels $\delta f=\left\langle f^2-\left\langle f\right\rangle^2\right\rangle^{1/2}$ and skewnesses $\gamma f=\left\langle f^3-\left\langle f\right\rangle^3\right\rangle/\delta f^3$, where angular brackets denote an average over toroidal direction and time within the saturated phase. Our statistical analysis was robust against averaging over different time windows. In fig.~\ref{fig:profiles_circular} profiles, fluctuation levels and skewnesses taken at outboard mid plane position are shown for density, electron temperature and pressure. There is a kink in the density and pressure at the last closed flux surface and a little distance outside also for the temperature. The electrostatic potential follows in the SOL $\phi\approx\Lambda T_e$ and deviates from this in the closed field line region where the potential at the inner limiting flux surface is prescribed as $\left.\phi\right|_{\rho_{min}}=\left\langle\Lambda T_e\right\rangle_{\rho=1}$. The absolute fluctuation level for the density in the closed field line region is below $0.13$ and still around $0.05$ in the near SOL. This translates into relative fluctuation levels ($\delta f / \left\langle f\right\rangle$) of around $30-40\%$, which substantiates again the importance of a global model. The skewnesses of the density and pressure approach zero in the region of its strongest gradients, which is an indication for a Gaussian probability density function and uncorrelated turbulence, which is driven in this region. Towards the core the skewnesses become negative which is an indication for holes and it becomes positive in the SOL indicating the presence of blobs. 

\subsection{Comparison with diverted geometry}\label{subsec:divturb}
\begin{figure}
\includegraphics[trim=0 0 0 0,clip,width=1.0\linewidth]{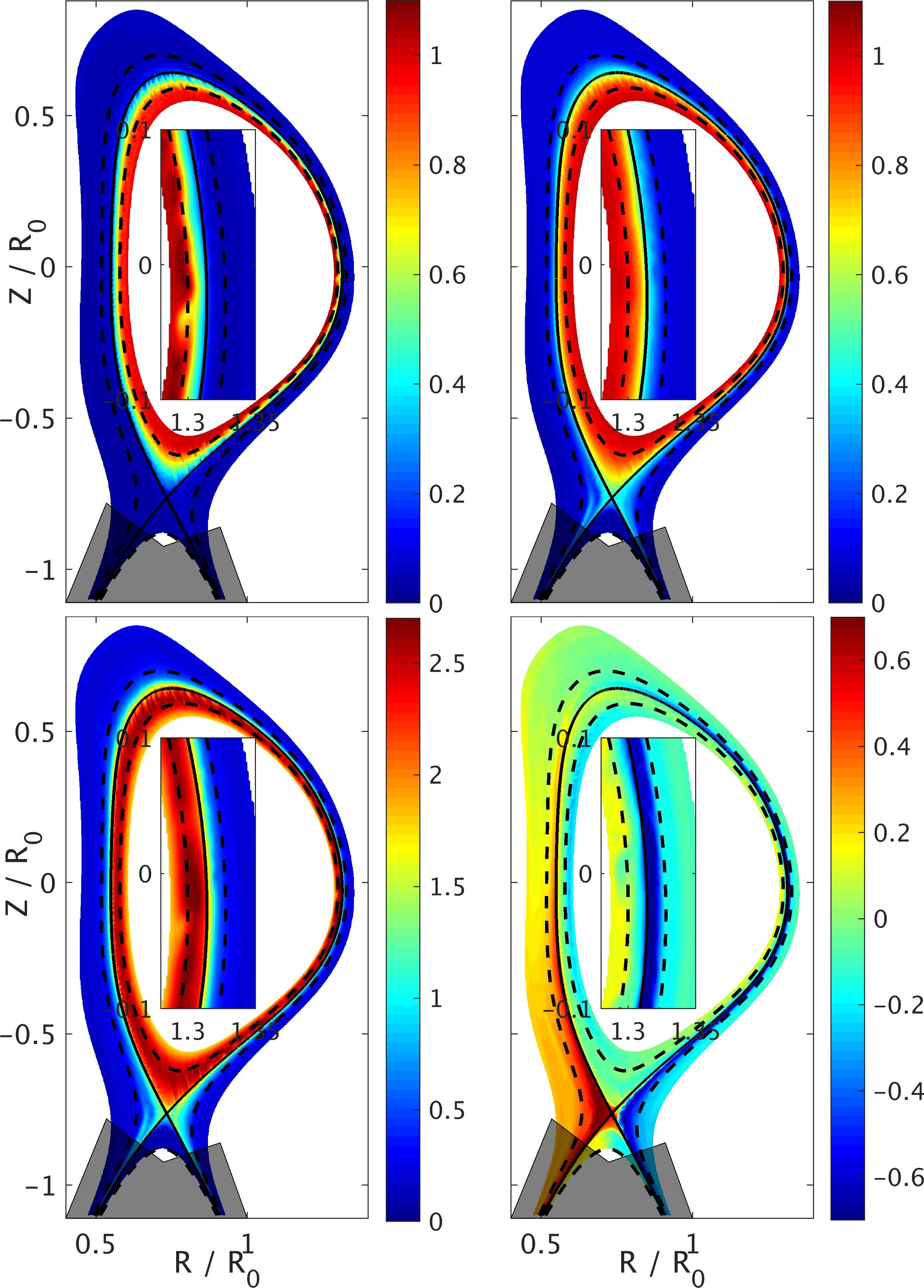}
\caption{Snapshots of density (top left), electron temperature (top right), electrostatic potential (bottom left) and parallel ion velocity (bottom right) in diverted geometry at $t=48.25$. Insets show outboard mid plane region enlarged. Solid black line indicates separatrix, dashed black lines flux surfaces $\rho=0.95$ and $\rho=1.05$ as reference. Gray shaded area marks penalization region due to divertor plates.}
\label{fig:xsnaps}
\end{figure}

Snapshots for the diverted case are shown in fig.~\ref{fig:xsnaps}, where again a clear distinction between closed field line region, SOL and private flux region in the dynamics is obvious. From the snapshots there seems to be a qualitative difference with respect to circular geometry: The turbulence in the saturated state is generally more quiescent especially in the SOL and we do not identify blobs at outboard mid plane as clearly as in circular geometry.

\begin{figure}
\includegraphics[trim=0 0 0 0, clip,width=1.0\linewidth]{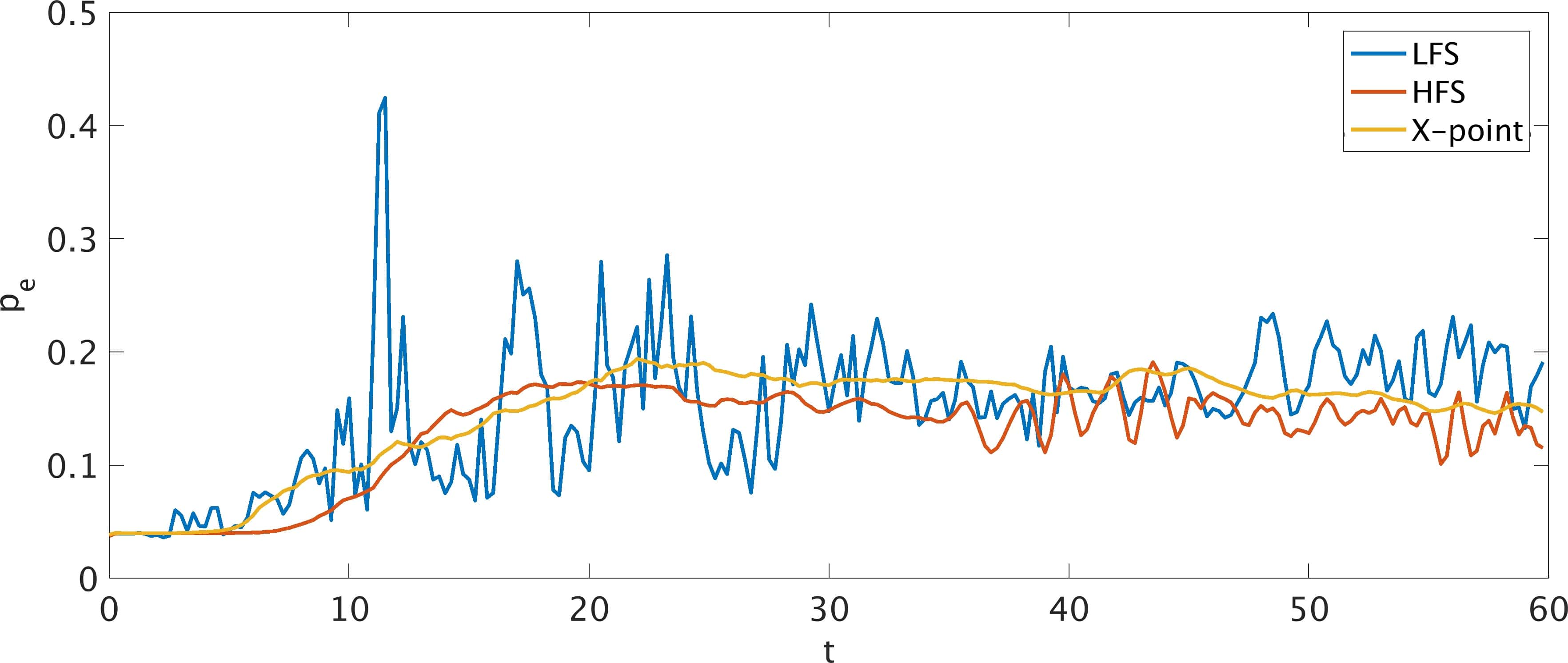}
\caption{Time traces of pressure on the separatrix at different poloidal positions for diverted geometry.}
\label{fig:xtraces}
\end{figure}

Firstly, we consider again time traces of pressure at different poloidal positions on the separatrix in fig.~\ref{fig:xtraces}. In comparison to circular geometry (see fig.~\ref{fig:ctraces}) the dynamics is generally more quiescent with smaller fluctuations, poloidal asymmetries are much stronger pronounced and the course into saturation is more complex and takes much longer. Until $t\approx30$ the turbulence exhibits a strong ballooning character with violent fluctuations at the LFS, whereas the HFS is rather quiescent. After $t\approx30$ the fluctuation levels approach each other whereas the signal at the X-point remains always rather quiescent. 

\begin{figure}
\includegraphics[trim=0 0 0 0,clip,width=1.0\linewidth]{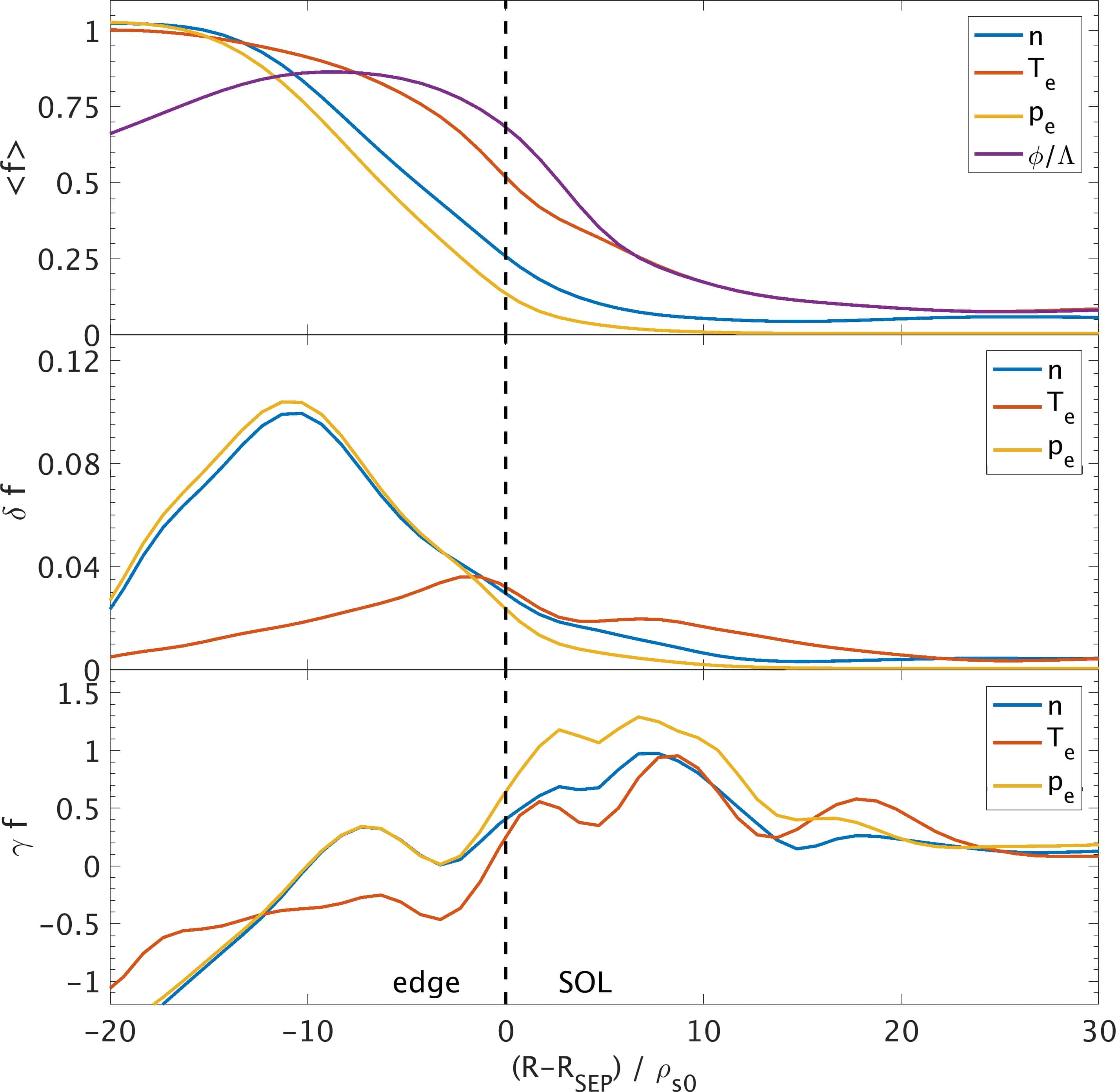}
\caption{Profiles (top), fluctuation levels (center) and skewnesses (bottom) for diverted geometry cut through outboard mid plane position (Note the different range of the x-axis in comparison to fig.~\ref{fig:profiles_circular}).}
\label{fig:profiles_diverted}
\end{figure}

Profiles, fluctuation levels and skewnesses for the diverted simulation are shown in fig,~\ref{fig:profiles_diverted}. Compared to the circular case (see fig.~\ref{fig:profiles_circular}), the profiles are steeper and the SOL-width is smaller such that the pressure in the SOL is lower. The fluctuation levels, especially in the SOL, are also reduced with a relative level of around $15-25\%$. The skewnesses are qualitatively similar indicating that blobs in the SOL are still present, although it might not directly be visible from the snapshots in fig.~\ref{fig:xsnaps}.

\begin{figure}
\centering
\includegraphics[trim=0 0 0 0,clip,width=1.0\linewidth]{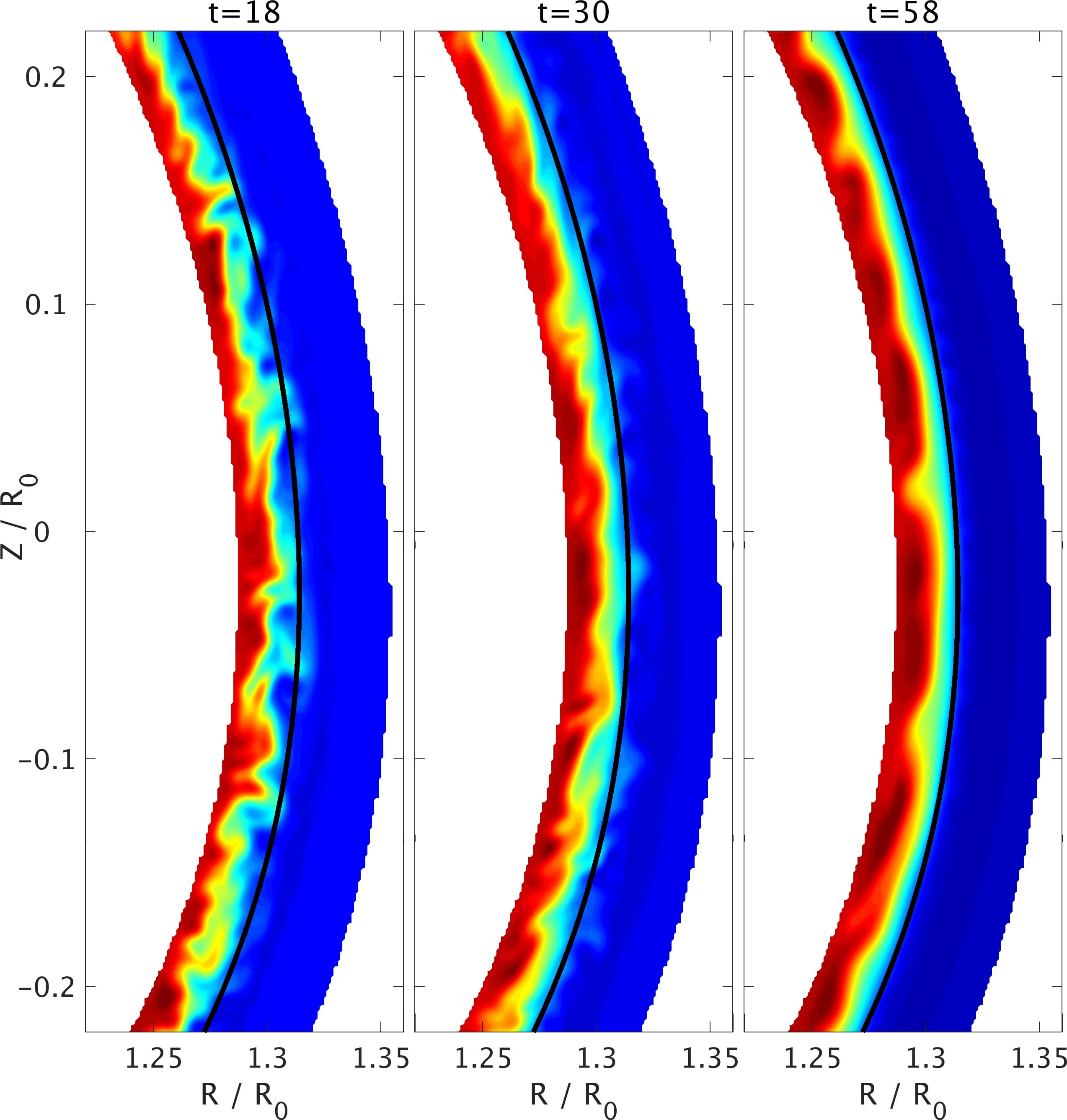}\\
\includegraphics[trim=0 0 0 0,clip,width=0.6\linewidth]{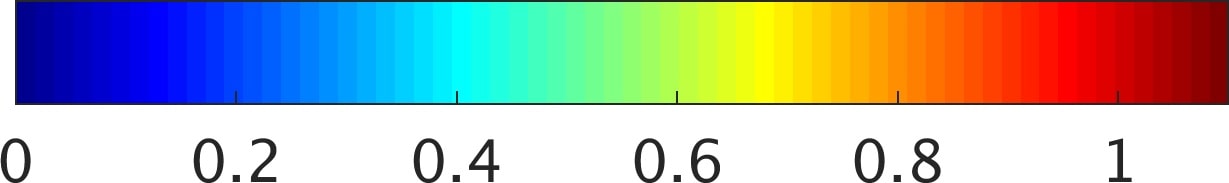}
\caption{Snapshots of density in outboard mid plane region for diverted geometry at time $t=18$ (left), $t=30$ (center) and $t=58$ (right). For temporal orientation of the simulation see fig.~\ref{fig:xtraces}.}
\label{fig:xsnaps_series}
\end{figure}

\begin{figure}
\centering
\includegraphics[trim=0 0 0 0,clip,width=1.0\linewidth]{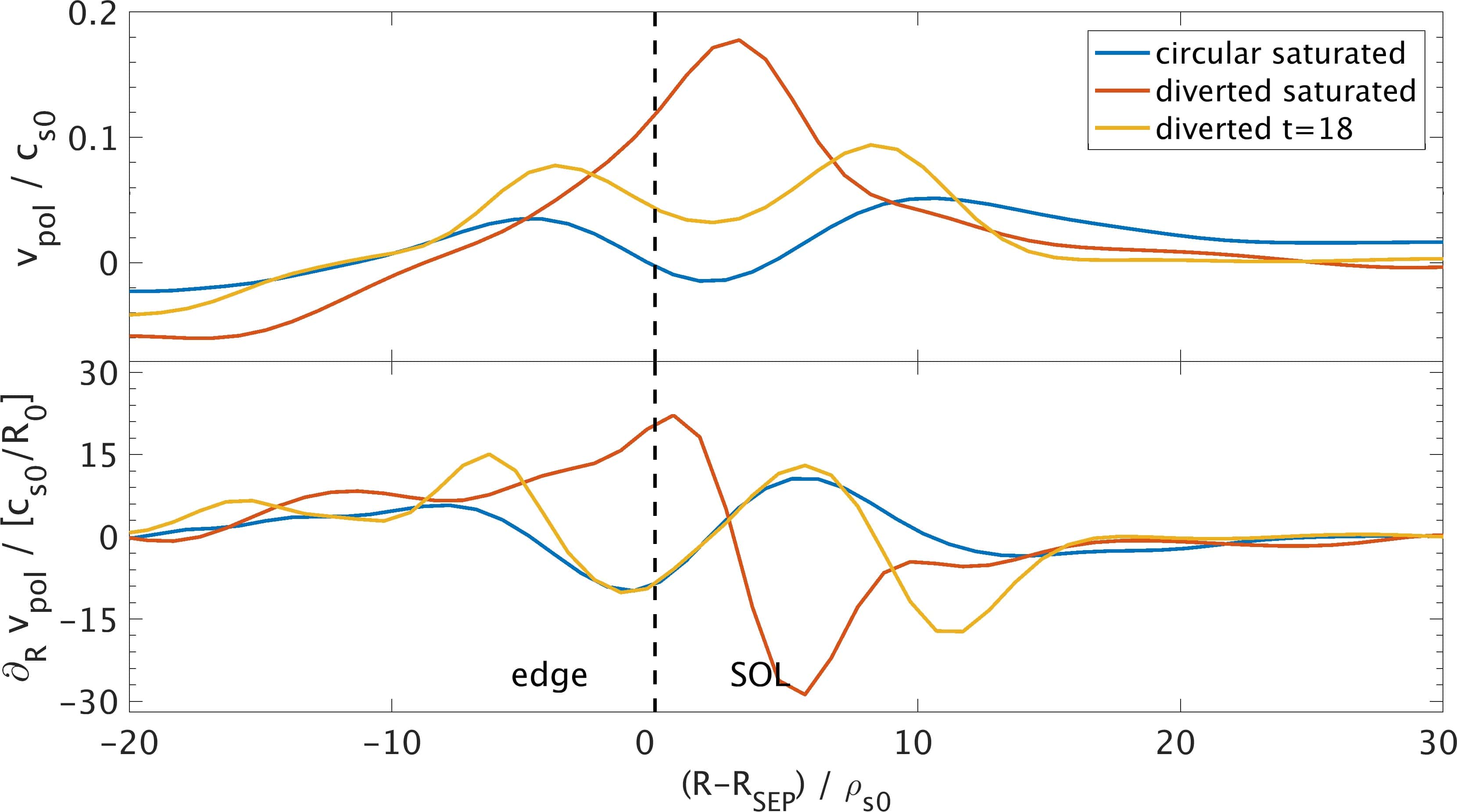}
\caption{Poloidal flow velocity (top) and flow shear (bottom) at outboard mid plane. Whereas the flow profiles of the circular geometry differs qualitatively from the diverted geometry in saturated state, it is very similar to the diverted flow profile at time $t=18$.}
\label{fig:flow_profiles}
\end{figure}

To illustrate the course of the diverted simulation into its saturated state we show a series of density snapshots in the outboard mid plane region in fig.~\ref{fig:xsnaps_series} at different times during the simulation. At an early time ($t=18$) \--but still well after the initial onset of turbulence\-- there are strong fluctuations and blobs can clearly be observed. The fluctuations reduce gradually and slowly until in the saturated state the dynamics becomes relatively quiescent. This transition and the qualitative difference to the circular geometry correlates with the rise of a poloidal shear flow illustrated in fig.~\ref{fig:flow_profiles}. When the flow profile of the diverted geometry at $t=18$ resembles the saturated flow profile in circular geometry the turbulence looks similar (Compare fig.~\ref{fig:xsnaps_series} left with fig.~\ref{fig:csnaps} upper left inset). In the saturated state, a strong poloidal flow is built up exhibiting a slightly larger and qualitatively different shearing rate that swamps away turbulent fluctuations. However, we have to stress here again that that the radial electric field and therefore the flow profile in this case is probably not very realistic as the ion temperature dynamics, which has a significant impact, is not yet included in the model and the core boundary condition for the potential is also chosen somewhat ad-hoc. Besides revealing the need for an accurate and self-consistent description for the radial electric field, the study shows that geometry has a significant qualitative effect on turbulence.

\begin{figure}
a)\\
\includegraphics[trim=0 0 0 0, clip,width=0.7\linewidth]{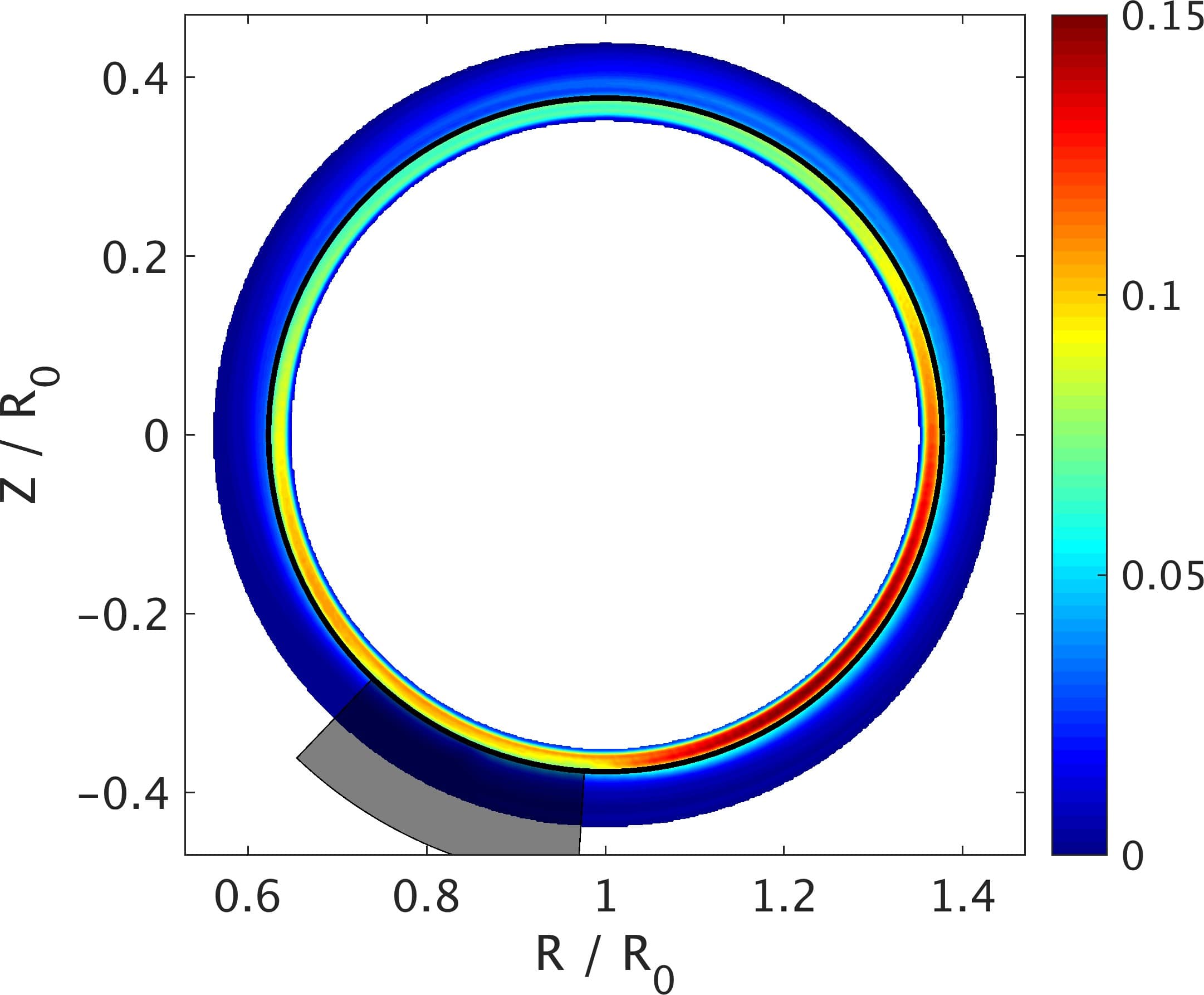}\\
b)\\
\includegraphics[trim=0 0 0 0, clip,width=0.7\linewidth]{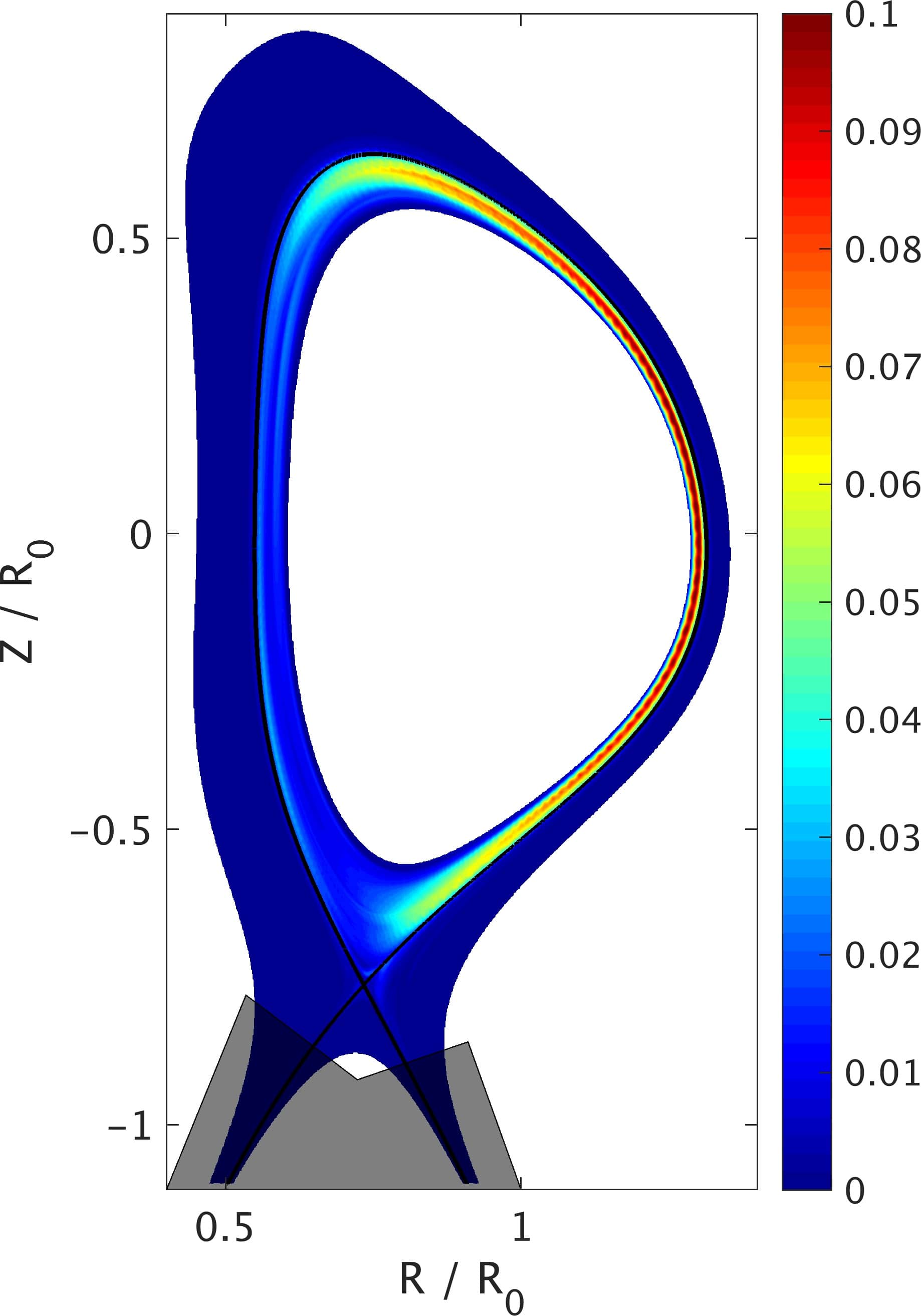}
\caption{Pressure fluctuation level $\delta p_e$ for a) circular geometry and b) diverted geometry.}
\label{fig:snaps_perms}
\end{figure}

Another qualitative difference between circular and diverted geometry are poloidal asymmetries. Whereas the fluctuation level in circular geometry varies within a flux surface only by a factor of two it varies in diverted geometry by an order of magnitude (see fig.~\ref{fig:snaps_perms}). Approaching the X-point region from LFS there is a strong drop in the fluctuation level. The poloidal asymmetry can also be seen by comparing again the time traces in figs.~\ref{fig:ctraces} and \ref{fig:xtraces}. Especially during earlier phases of the diverted simulation the fluctuations seem to be concentrated at the LFS, whereas the HFS and especially the X-point region are rather quiescent.

\begin{figure}
\includegraphics[trim=0 0 0 0,clip,width=1.0\linewidth]{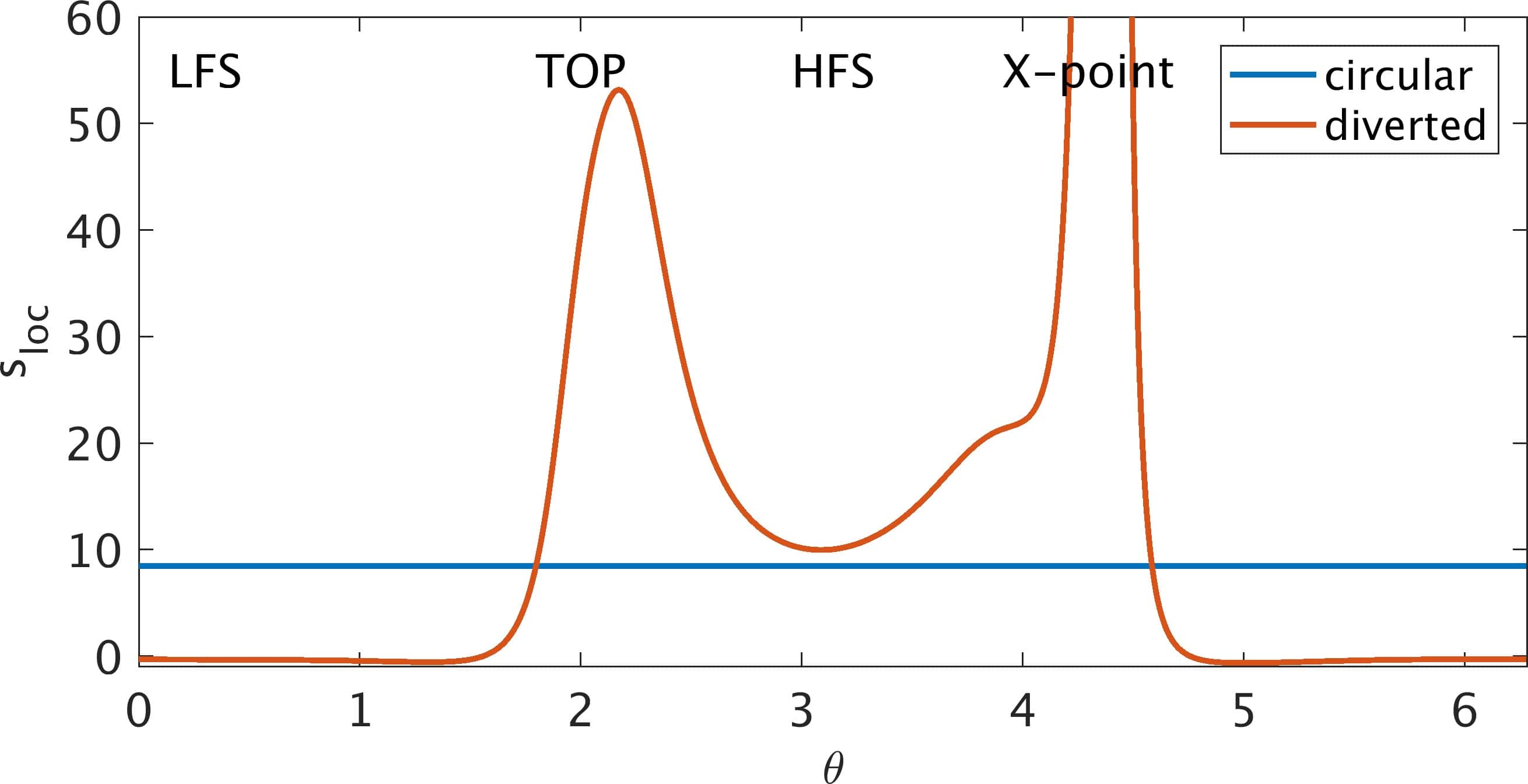}
\caption{Local magnetic shear $s_{loc}$ on flux surface $\rho=0.992$ close to separatrix for circular and diverted geometry as function of geometric poloidal angle.}
\label{fig:shear_local}
\end{figure}

To give a reason for the stronger poloidal asymmetries in diverted geometry we investigate the effect of magnetic geometry on turbulent fluctuations and consider the local magnetic shear:
\begin{align*}
s_{loc}(\theta,\rho_N)=\frac{\partial}{\partial\rho}\left(\frac{B^\varphi}{B^\theta}\right),
\end{align*}
where $\tan\theta:=Z/(R-R_0)$ is the geometric poloidal angle. A plot of the local magnetic shear for flux surfaces just inside the separatrix is shown in fig.~\ref{fig:shear_local}. Whereas the local shear is obviously constant in circular geometry, it follows a complicated course in diverted geometry, i.e.~it is very low in the outboard mid plane region, increases towards the top and high field side region and approaches a singularity at the X-point. Turbulent structures, which are driven in the outboard mid plane region due to unfavourable curvature, become distorted in the perpendicular direction due to local magnetic shear (see also \cite{manz:reyn18}). Being strongly distorted, i.e.~especially in the vicinity of the X-point, the fluctuations are damped due to perpendicular dissipation. Therefore the X-point ultimately acts as kind of barrier for fluctuations \cite{farina:xpoint93} (See also resistive X-point mode by Myra et al.~\cite{myra:resisitivexpointmode00}). This explains the drop of the fluctuation level near the X-point towards HFS in our simulation (see fig.~\ref{fig:snaps_perms}b). There is also a drop in fluctuation level in the top region towards HFS, where local magnetic shear is also relatively large.

\begin{figure}
\includegraphics[trim=0 0 0 0,clip,width=0.8\linewidth]{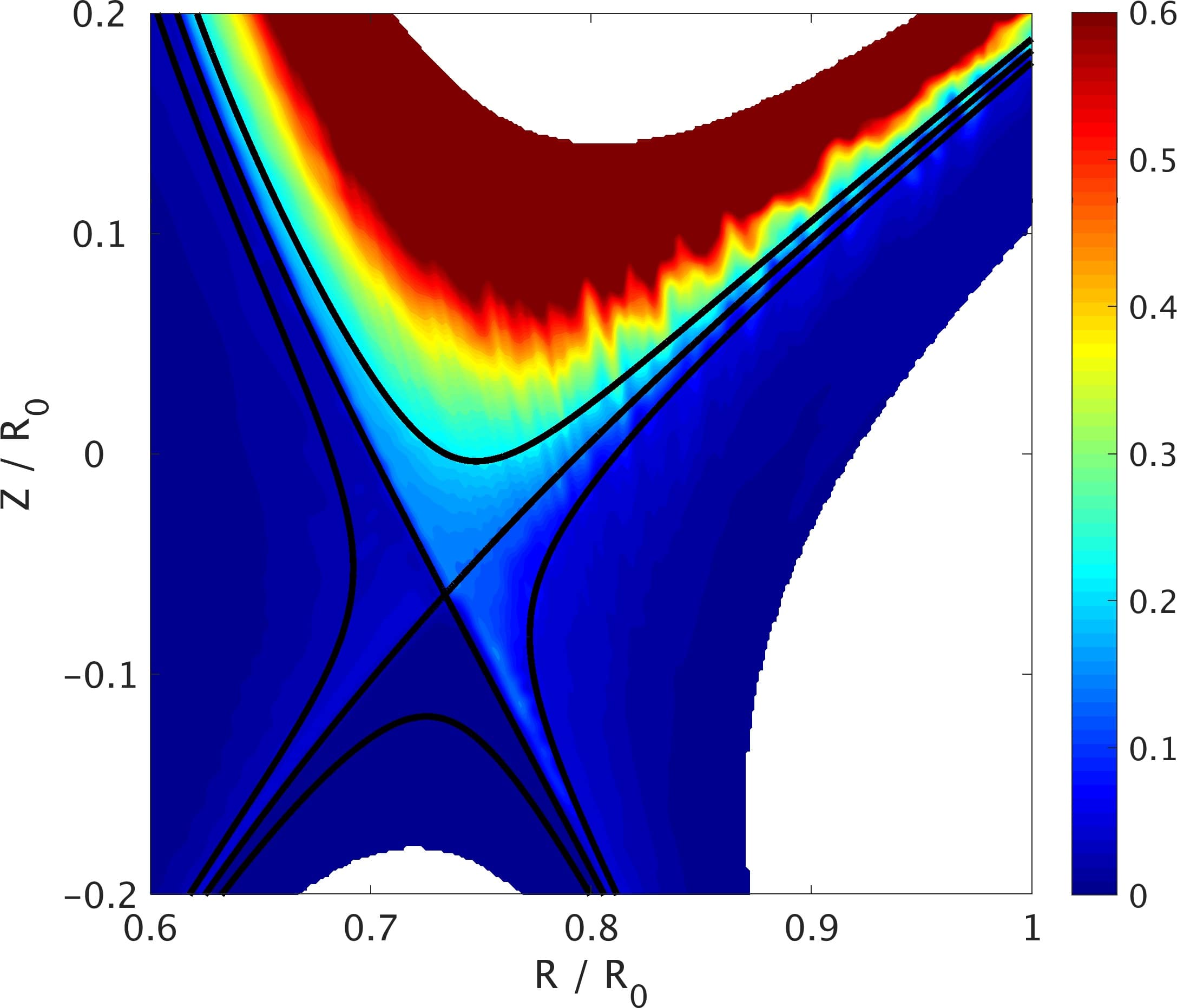} 
\caption{Snapshot of pressure at $t=30$ where the distortion of structures towards the X-point is visible. Solid lines show $\rho=0.99,\,1.0,\,1.01$ flux surfaces, illustrating flux expansion.}
\label{fig:xpoint_detailed}
\end{figure}

In contrast to flux-aligned approaches GRILLIX does not suffer from coordinate singularity and a loss of resolution near the X-point due to flux expansion, but the FCI approach allows an accurate treatment of the dynamics around the X-point. A detailed view of the pressure at the X-point is shown in fig.~\ref{fig:xpoint_detailed}. Fluctuations approaching the X-point fan out radially becoming ever narrower in the poloidal direction, which illustrates the mechanism described in the previous paragraph.

\subsection{Comparison to Boussinesq system}
\label{subsec:bsqcomparison}
Although the Boussinesq approximation is factually not justified in the edge and SOL, it has often been employed in various codes and in various forms for numerical reasons (e.g.~\cite{stegmeir:ppcf18,paruta:gbsx18,dudson:hermes17,tamain:tokam3x16}). For that matter the effect of the Boussinesq approximation has been a long standing discussion, and since recently there have been several developments on its abolishment \cite{halpern:gbs16,zhu:gdb18}. At the same time the impact of the Boussinesq approximation has mainly been studied at isolated phenomena, i.e.~blobs \cite{halpern:gbs16,angus:nonbsq14,militello:blobs17}, whereas we investigate here the effect of the Boussinesq approximation in a fully turbulent environment. For this purpose we modify the vorticity equation (\ref{eq:vorticity}) to:
\begin{align}
\frac{d_u}{dt}\Omega_{bsq}=&-nT_e\left[\mathcal{C}(\theta_n+\xi_e)\right]+\nabla\cdot\left(\mathbf{b}j_\parallel\right) + \mathcal{D}_w\left(\Omega_{bsq}\right),
\label{eq:vorticity_boussinesq}
\end{align}
where we define the Boussinesq vorticity as $\Omega_{bsq}=\nabla\cdot\left(\frac{1}{B^2}\nabla_\perp\phi\right)$, i.e.~we drop the spatio-temporal dependence of the density in the polarisation term completely. We note that we also tried another form for the Boussinesq approximation, where the density is taken out of the divergence with retention of its spatio-temporal dependence, i.e.~$\Omega_{bsq(alt)}=n\nabla\cdot\left(\frac{1}{B^2}\nabla_\perp\phi\right)$. This alternative form breaks the conservation property of the quasi-neutrality equation, for which reason we observed a strong spurious poloidal flow rising that eventually caused our simulations to crash \cite{ross:phd18,ross:bossinesq18}.
  
\begin{figure}
\includegraphics[trim=0 0 0 0,clip,width=1.0\linewidth]{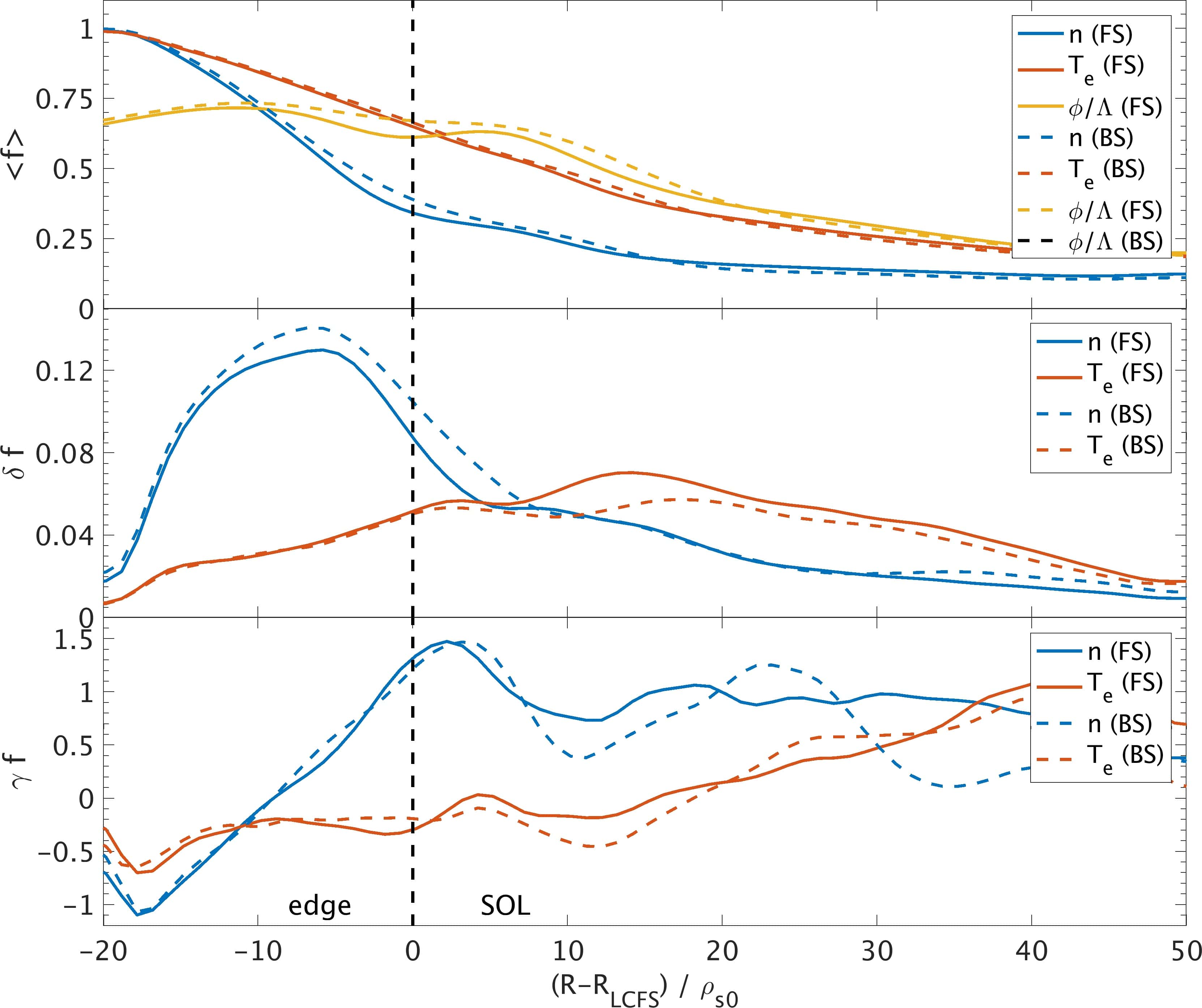} 
\caption{Comparison of global model (FS, solid lines) and Boussinesq modified model (BS, dashed lines). Profiles (top), fluctuation level (center) and skewnesses (bottom) at outboard mid plane are shown.}
\label{fig:bsq_comparison}
\end{figure}

Based on the circular limited case, we directly compare our simulation results from section \ref{subsec:circturb}, obtained with the global model, to the Boussinesq reduced model. We restarted the circular case from scratch, and except from the Boussinesq modification there are no other modifications neither in physical nor numerical parameters. The result of this comparison is shown in fig.~\ref{fig:bsq_comparison}, i.e.~profiles, fluctuation levels and skewnesses at outboard mid plane. Our results were again checked to be robust against statistical noise via averaging over distinct time windows within the saturated phase. There are only small quantitative differences between the global and the Boussinesq model. There is basically no difference in the temperature profile and only a small difference of around $10\%$ in the density profile around the separatrix. The fluctuation levels differ at most around $10\%$ and the skewnesses are very similar in both cases. In conclusion, the  Boussinesq approximation has only minor quantitative effects for the case considered here.

\subsection{Convergence analysis and impact of heat conductivity}\label{subsec:convergenc_and_heatconductivity}
A posteriori, we subject our results to a convergence check and study the impact of heat conductivity, as we decreased it artificially. We perform this study only for circular geometry as it is computationally generally less expensive than the diverted case and its saturation time is shorter.

\begin{figure}
\includegraphics[trim=0 0 0 0,clip,width=1.0\linewidth]{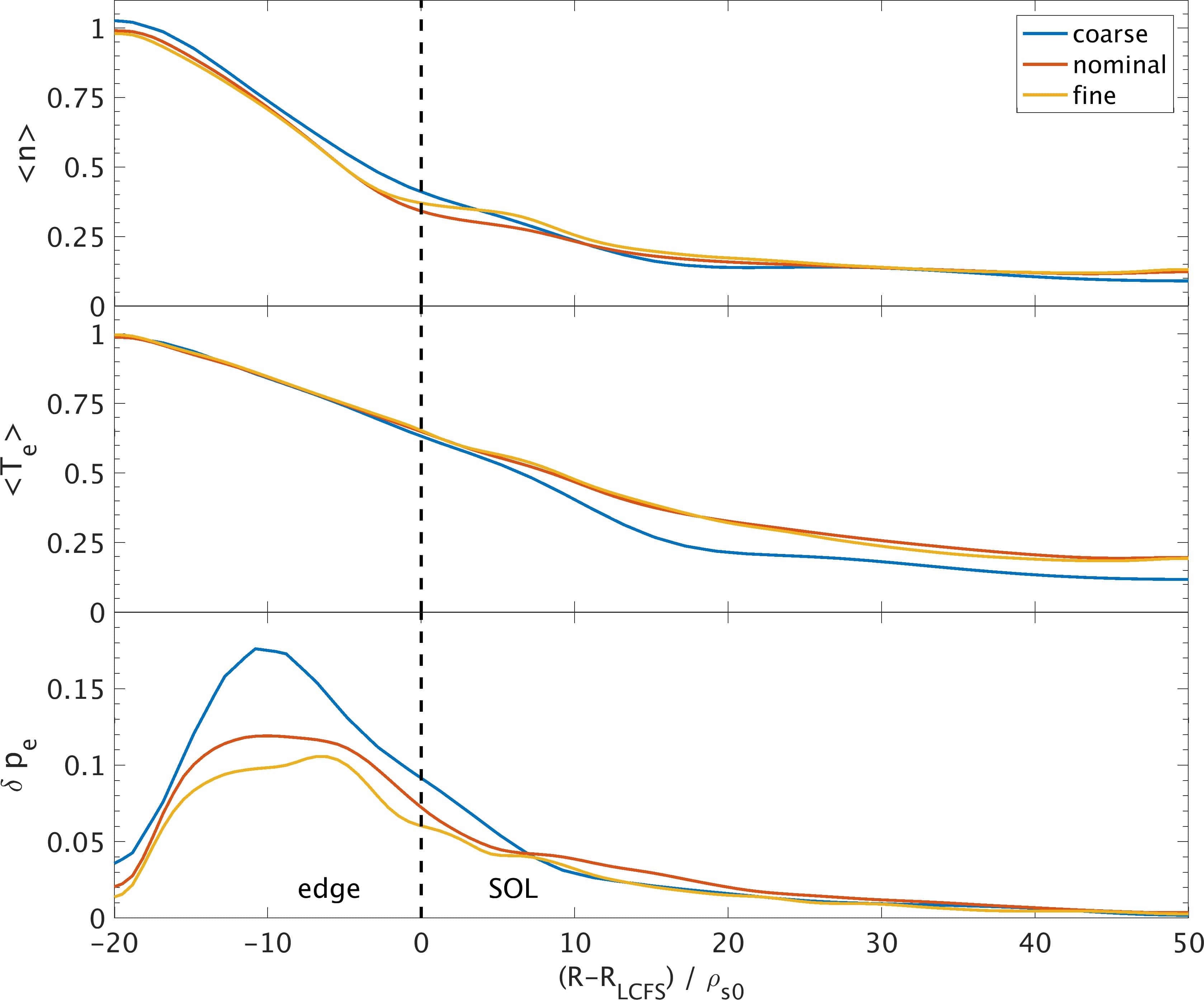}
\caption{Result of convergence analysis: Outboard mid plane profiles for density (top), electron temperature (center) and pressure fluctuation level (bottom) obtained with coarse, nominal and fine resolution.}
\label{fig:convergence_check}
\end{figure}

For a convergence check we ran the circular case at nominal resolution, at half resolution and at a resolution that was increased by a factor of one third. We correspondingly adapted also other numerical parameters, i.e.~we decreased numerical dissipation coefficients with increasing resolution. Moreover, we also varied the poloidal and radial decay lengths of the penalisation functions $\chi$ in order to exclude spurious effects from our penalisation approach. The numerical parameters employed are listed explicitly in table \ref{table:convergence_check}. The obtained profiles and pressure fluctuation level at LFS are shown in fig.~\ref{fig:convergence_check}. Whereas there is a deviation in the temperature profile to the coarse resolution case, the fine and nominal resolution match here very well. The density profiles between nominal and fine resolution match overall well apart from the density shoulder at the last closed flux surface, which is slightly more pronounced at finer resolution. Also the pressure fluctuation converges and deviation between nominal and fine resolution could also be owed to the fact that for the fine case not as much statistics as for the other cases was available due to computational constraints. 

\begin{table}
\centering
\begin{tabular}{l|ccccccc}
& $h_\perp$ & $N_{pol}$ &  $\nu$ & $\mu$ & $w_\theta$ & $w_\rho$\\
\hline
coarse  & $2.0$  & $16$  & $500.0$  & $5.0\cdot10^{-2}$ & $0.1$ & $6.0\cdot10^{-3}$ \\
nominal & $1.0$  & $32$  & $10.0 $  & $2.5\cdot10^{-2}$ & $7.5\cdot10^{-2}$ & $4.0\cdot10^{-3}$ \\
fine    & $0.67$ & $48$  & $2.0  $  & $1.5\cdot10^{-2}$ & $5.0\cdot10^{-2}$ & $3.0\cdot10^{-3}$
\end{tabular}
\caption{Numerical parameters used for convergence check. $w_\theta$ is poloidal decay length in radians and $w_\rho$ radial decay length in units of $R_0$ for penalisation function, which is parametrized via $\tanh$ functions.}
\label{table:convergence_check}
\end{table}

\begin{figure}
\includegraphics[trim=0 0 0 0,clip,width=1.0\linewidth]{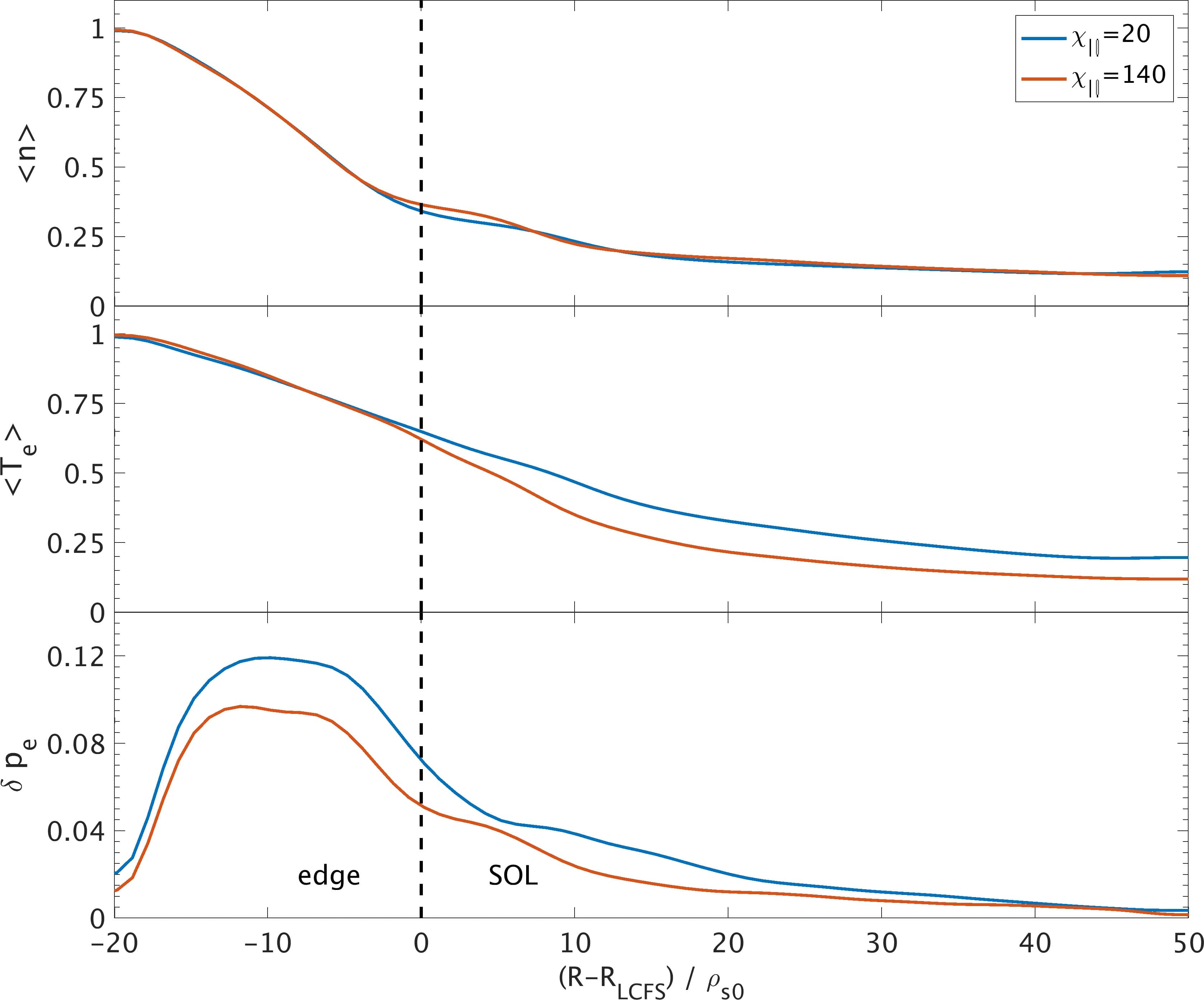}
\caption{Impact of heat conductivity on density (top) temperature profile (center) and pressure fluctuation level (bottom) at outboard mid plane position.}
\label{fig:heat_conductivityimpact}
\end{figure}

Compared to realistic COMPASS parameters we employed a significantly reduced artificial heat conductivity of $\chi_{\parallel0}=20$ instead of $\chi_{\parallel0}=340$ and correspondingly also decreased the effective sheath transmission factor to $\gamma_e=0.15$ instead of $\gamma_e=2.5$. In order to study the effect of this we restarted with reduced time step the circular limited simulation from a saturated state with a more realistic heat conductivity of $\chi_{\parallel0}=140$ and $\gamma_e=1.0$. The obtained profiles and pressure fluctuation level are shown in fig.~\ref{fig:heat_conductivityimpact}. While the increased heat conductivity does not alter the density profile significantly it steepens the temperature profile in the SOL. The fluctuation level reduces slightly which is consistent with \cite{zhu:gdb17}, where a similar study was carried out.

\section{Conclusions and Outlook}\label{sec:Conclusions_and_outlook}
By incorporating full parametric dependencies and relaxing the Boussinesq approximation, GRILLIX was extended to a global fluid turbulence code for the tokamak edge and SOL, i.e.~no assumption about fluctuation amplitudes of density or temperature is made. Further new features are electromagnetic and electron-thermal dynamics, and the implementation of the extended model was verified by analytical means and the Method of Manufactured Solutions (MMS). The flux-coordinate independent approach (FCI) is employed in GRILLIX in order to deal with realistic geometries avoiding coordinate singularities at the X-point or separatrix. As boundary contours do in general not conform with the computational grid nor the magnetic field as exceptional direction, a penalisation technique is used in order to treat boundary conditions at the target plates.

Turbulence simulations in circular geometry with toroidal limiter and in realistic diverted geometry at otherwise comparable parameters were presented, where parameters were chosen being characteristic for the COMPASS tokamak. A relative fluctuation level of around $30-40\%$ in the SOL, with isolated highly intermittent phenomena of up to $200\%$ fluctuation level were observed in circular geometry, which implies that a global description for SOL turbulence is indeed important. At the same time a direct comparison to a Boussinesq reduced model revealed that the Boussinesq approximation had only minor quantitative effects. However, this conclusion might possibly not hold true for other parameter regimes \cite{ross:bossinesq18} and therefore it is safest to abolish it consequently. Besides this, other global features, e.g.~parametric dependencies of the heat conductivities and resistivity, might play a more important role. In diverted geometry the turbulence was found to be generally more quiescent in the saturated phase, which is related to a different zonal flow structure. Moreover, the diverted geometry exhibits stronger poloidal asymmetries that can be attributed to local magnetic shear, which is fundamentally different in the edge region between circular and diverted geometry. Approaching regions of strong local magnetic shear turbulent structures become distorted towards the high field side and subsequently subject to enhanced perpendicular dissipation. As this mechanism is very strong near the X-point, it tends to disconnect low field side and high field side, where curvature is favourable \cite{farina:xpoint93}. In conclusion the presented studies point out the importance of global effects due to the presence of large intermittent fluctuations and qualitative differences between circular and diverted geometry.  

However, the results have to be considered qualitative as there are yet important effects in the physical modelling missing, i.e.~ion thermal dynamics, interaction with neutrals and more realistic sheath boundary conditions that could also take into account glancing angles of incidence. Our study revealed also the importance of the radial electric field, and its self-consistent description requires either an adjustment of the core boundary conditions or simply a complete abolishment of the core boundary by performing computationally intense full tokamak simulations. Besides this, there are also numerical constraints in order to achieve realistic parameter regimes concerning computational efficiency. The strongest time step limitation arises from the parallel non-linear heat conduction scaling strongly with $\chi_\parallel\propto T_e^{5/2}$, for which an implicit treatment will be necessary. Efforts to extend GRILLIX in these directions are currently ongoing.

In the near future our studies  would concentrate on investigation of advanced divertor concepts such as double-null, snowflake or super-X configurations, whose treatment is straight forward with GRILLIX as its numerics is independent of flux surfaces.

\section{Acknowledgements}
The authors would like to thank A.~Bottino for computational support. This work has been carried out within the framework of the EUROfusion Consortium and has received funding from the Euratom research and training program 2014-2018 and 2019-2020 under grant agreement No 633053. The views and opinions expressed herein do not necessarily reflect those of the European Commission.

\bibliography{stegmeir_pop19_rev.bbl}

\begin{thebibliography}{10}
\expandafter\ifx\csname url\endcsname\relax
  \def\url#1{\texttt{#1}}\fi
\expandafter\ifx\csname urlprefix\endcsname\relax\def\urlprefix{URL }\fi
\expandafter\ifx\csname href\endcsname\relax
  \def\href#1#2{#2} \def\path#1{#1}\fi

\bibitem{eich:solwidth13}
T.~Eich, A.~W. Leonard, R.~A. Pitts, W.~Fundamenski, R.~J. Goldston, T.~Gray,
  A.~Herrmann, A.~Kirk, A.~Kallenbach, O.~Kardaun, A.~S. Kukushkin,
  B.~LaBombard, R.~Maingi, M.~A. Makowski, A.~Scarabosio, B.~Sieglin, J.~Terry,
  A.~Thornton, A.~U. Team, J.~E. Contributors, Nucl. Fusion 53 (2013) 093031.

\bibitem{halpern:solwidth13}
F.~Halpern, P.~Ricci, B.~Labit, I.~Furno, S.~Jolliet, J.~Loizu, A.~Mosetto,
  G.~Arnoux, J.~Gunn, J.~Horacek, M.~Ko{\v c}an, B.~LaBombard, C.~Silva, J.-E.
  Contributors, Nucl. Fusion 53 (2013) 122001.

\bibitem{wiesen:solps15}
S.~Wiesen, D.~Reiter, V.~Kotov, M.~Baelmans, W.~Dekeyser, A.~Kukushkin,
  S.~Lisgo, R.~Pitts, V.~Rozhansky, G.~Saibene, I.~Veselova, S.~Voskoboynikov,
  J. Nucl. Mater. 463 (2015) 480.

\bibitem{braginskii65}
S.~I. Braginskii, Transport processes in a plasma, in: A.~M.~A. Leontovich
  (Ed.), Reviews of Plasma Physics, Vol.~1, Consultants Bureau, 1965.

\bibitem{ricci:gbs12}
P.~Ricci, F.~D. Halpern, S.~Jolliet, J.~Loizu, A.~Mosetto, A.~Fasoli, I.~Furno,
  C.~Theiler, Plasma Phys. Contr. F. 54 (2012) 124047.

\bibitem{halpern:gbs16}
F.~D. Halpern, P.~Ricci, S.~Jolliet, J.~Loizu, J.~Morales, A.~Mosetto,
  F.~Musil, F.~Riva, T.~M. Tran, C.~Wersal, J. Comput. Phys. 315 (2016) 388.

\bibitem{paruta:gbsx18}
P.~Paruta, P.~Ricci, F.~Riva, C.~Wersal, C.~Beadle, B.~Frei, Phys. Plasmas 25
  (2018) 112301.

\bibitem{dudson:hermes17}
B.~D. Dudson, J.~Leddy, Plasma Phys. Contr. F. 59 (2017) 054010.

\bibitem{tamain:tokam3x16}
P.~Tamain, H.~Bufferand, G.~Ciraolo, C.~Colin, D.~Galassi, P.~Ghendrih,
  F.~Schwander, E.~Serre, J. Comput. Phys. 321 (2016) 606.

\bibitem{zhu:gdb18}
B.~Zhu, M.~Francisquez, B.~N. Rogers, Comput. Phys. Commun. 232 (2018) 46.

\bibitem{stegmeir:ppcf18}
A.~Stegmeir, D.~Coster, A.~Ross, O.~Maj, K.~Lackner, E.~Poli, Plasma Phys.
  Contr. F. 60 (2018) 035005.

\bibitem{hariri:fenicia13}
F.~Hariri, M.~Ottaviani, Comput. Phys. Commun. 184 (2013) 2419.

\bibitem{stegmeir:cpc16}
A.~Stegmeir, D.~Coster, O.~Maj, K.~Hallatschek, K.~Lackner, Comput. Phys.
  Commun. 198 (2016) 139.

\bibitem{stegmeir:fciaddendum17}
A.~Stegmeir, O.~Maj, D.~Coster, K.~Lackner, M.~Held, M.~Wiesenberger, Comput.
  Phys. Commun. 213 (2017) 111.

\bibitem{wootton:edgesolcharacter90}
A.~J. Wootton, B.~A. Carreras, H.~Matsumoto, K.~McGuire, W.~A. Peebles, C.~P.
  Ritz, P.~W. Terry, S.~J. Zweben, Phys. Fluids. B 2 (1990) 2879.

\bibitem{angus:nonbsq14}
J.~R. Angus, S.~I. Krasheninnikov, Phys. Plasmas 21 (2014) 112504.

\bibitem{militello:blobs17}
F.~Militello, B.~Dudson, L.~Easy, A.~Kirk, P.~Naylor, Plasma Phys. Contr. F. 59
  (2017) 125013.

\bibitem{ross:phd18}
A.~Ross, {Extension of GRILLIX: Towards a global fluid turbulence code for
  realistic magnetic geometries}, Ph.D. thesis, Technichal University of Munich
  (2018).

\bibitem{ross:bossinesq18}
A.~Ross, A.~Stegmeir, P.~Manz, D.~Coster, W.~Zholobenko, {On the nature of blob
  propagation and generation in Large Plasma Device: Global GRILLIX studies},
  submitted to Phys. Plasmas.

\bibitem{panek:compass15}
R.~P{\'a}nek, J.~Ad{\'a}mek, M.~Aftanas, P.~B{\'i}lkov{\'a}, P.~B{\"o}hm,
  F.~Brochard, P.~Cahyna, J.~Cavalier, R.~Dejarnac, M.~Dimitrova1, O.~Grover,
  J.~Harrison, P.~H{\'a}{\v c}ek, J.~Havl{\'i}{\v c}ek, A.~Havr{\'a}nek,
  J.~Hor{\'a}{\v c}ek, M.~Hron, M.~Imr{\'i}{\v s}ek, F.~Janky, A.~Kirk,
  M.~Komm, K.~Kova{\v r}{\'i}k, J.~Krbec, L.~Kripner, T.~Markovi{\v c},
  K.~Mito{\v s}inkov{\'a}, J.~Mlyn{\'a}{\v r}, D.~Naydenkova, M.~Peterka,
  J.~Seidl, J.~St{\"o}ckel, E.~{\v S}tef{\'a}nikov{\'a}, M.~Tome{\v s},
  J.~Urban, P.~Vondr{\'a}{\v c}ek, M.~Varavin, J.~Varju, V.~Weinzettl,
  J.~Zajac, the COMPASS~team, Plasma Phys. Contr. F. 58 (2015) 014015.

\bibitem{farina:xpoint93}
D.~Farina, R.~Pozzoli, D.~D. Ryutov, Nucl. Fusion 33 (1993) 1315.

\bibitem{zeiler:drift_approx97}
A.~Zeiler, J.~F. Drake, B.~Rogers, Phys. Plasmas 4 (1997) 2134.

\bibitem{francisquez:phd18}
M.~Francisquez, {Global Braginskii modeling of magnetically confined boundary
  plasmas}, Ph.D. thesis, Dartmouth College, Hanover (New Hampshire) (2018).

\bibitem{salari:mms00}
K.~Salari, P.~Knupp, Code verification by the method of manufactured solutions,
  {Sandia National Laboratories}, {Sandia Report SAND2000-1444} (2000).

\bibitem{ross:nonbsq17}
A.~Ross, A.~Stegmeir, D.~Coster, Contrib. Plasm. Phys. 58 (2018) 478.

\bibitem{loizu:bndconds12}
J.~Loizu, P.~Ricci, F.~D. Halpern, S.~Jolliet, Phys. Plasmas 19 (2012) 122307.

\bibitem{stangeby:plasmaboundary00}
P.~C. Stangeby, {The Plasma Boundary of Magnetic Fusion Devices}, Plasma
  Physics Series, Institute of Physics Publishing, 2000.

\bibitem{krasheninnikov:blobs08}
S.~I. Krasheninnikov, D.~A. D'Ippolito, J.~R. Myra, J. Plasma Physics 74 (2008)
  679.

\bibitem{stroth:er11}
U.~Stroth, P.~Manz, M.~Ramisch, Plasma Phys. Contr. F. 53 (2011) 024006.

\bibitem{viezzer:er13}
E.~Viezzer, T.~P{\"u}tterich, G.~Conway, R.~Dux, T.~Happel, J.~Fuchs,
  R.~McDermott, F.~Ryter, B.~Sieglin, W.~Suttrop, M.~Willensdorfer, E.~Wolfrum,
  the ASDEX Upgrade~Team, Nucl. Fusion 53 (2013) 053005.

\bibitem{arakawa:scheme97}
A.~Arakawa, J. Comput. Phys. 135 (1997) 103.

\bibitem{hairer:dop85393}
E.~Hairer, S.~P. Norsett, G.~Wanner, {Solving ordinary Differential Equations
  I. Nonstiff Problems}, 2nd Edition, Springer Series in Computational
  Mathematics, Springer-Verlag, 1993.

\bibitem{hill:fci17}
P.~Hill, B.~Shanahan, B.~Dudson, Comput. Phys. Commun 213 (2017) 9.

\bibitem{shanahan:bsting18}
B.~Shanahan, B.~Dudson, P.~Hill, {Fluid simulations of plasma laments in
  stellarator geometries with BSTING}, arXiv (2018) 1808.08899v1.

\bibitem{shashkov:support95}
M.~Shashkov, S.~Steinberg, J. Comput. Phys. 118 (1995) 131.

\bibitem{shashkov:support96}
M.~Shashkov, {Conservative Finite-Difference Methods on General Grids}, CRC
  Press, 1996.

\bibitem{isoardi:penalisation10}
L.~Isoardi, G.~Chiavassa, G.~Ciraolo, P.~Haldenwang, E.~Serre, P.~Ghendrih,
  Y.~Sarazin, F.~Schwander, P.~Tamain, J. Comput. Phys. 229 (2010) 2220.

\bibitem{bufferand:edgesol2dpen13}
H.~Bufferand, B.~Bensiali, J.~Bucalossi, G.~Ciraolo, P.~Genesio, P.~Ghendrih,
  Y.~Marandet, A.~Paredes, F.~Schwander, E.~Serre, P.~Tamain, J. Nucl. Mater.
  438 (2013) S445.

\bibitem{karniadakis:bdf91}
G.~E. Karniadakis, M.~Israeli, S.~A. Orszag, J. Comput. Phys. 97 (1991) 414.

\bibitem{hackbusch:mgrid85}
W.~Hackbusch, {Multi-Grid Methods and Applications}, Springer-Verlag, 1985.

\bibitem{scott:habil00}
B.~Scott, {Low frequency fluid drift turbulence in magnetised plasmas},
  Habilitation thesis, Heinrich-Heine University D{\"u}sseldorf (2000).

\bibitem{scott:dwtimplicit88}
B.~D. Scott, J. Comput. Phys. 78 (1988) 114.

\bibitem{cerfon:equilibrium10}
A.~J. Cerfon, J.~P. Freidberg, Phys. Plasmas 17 (2010) 032502.

\bibitem{held:cpc16}
M.~Held, M.~Wiesenberger, A.~Stegmeir, Comput. Phys. Commun. 199 (2016) 29.

\bibitem{halpern:solballooning13}
F.~D. Halpern, S.~Jolliet, J.~Loizu, A.~Mosetto, P.~Ricci, Phys. Plasmas 20
  (2013) 052306.

\bibitem{zhu:gdb17}
B.~Zhu, M.~Francisquez, B.~N. Rogers, Phys. Plasmas 24 (2017) 055903.

\bibitem{manz:reyn18}
P.~Manz, A.~Stegmeir, B.~Schmid, T.~T. Ribeiro, G.~Birkenmeier, N.~Fedorczak,
  S.~Garland, K.~Hallatschek, M.~Ramisch, B.~D. Scott, Phys. Plasmas 25 (2018)
  072508.

\bibitem{myra:resisitivexpointmode00}
J.~R. Myra, D.~A. D'Ippolito, X.~Q. Xu, R.~H. Cohen, Phys. Plasmas 7 (2000)
  2290.

\end{thebibliography}

%
%
%
\end{document}